\documentclass[11pt]{JHEP3}

\let\ifpdf\relax
\usepackage{ifpdf}

\usepackage{graphicx}
\usepackage{dcolumn}
\usepackage{bm}
\usepackage{cite}
\usepackage{amsmath}
\usepackage{amssymb}

\title{Dark Matter in a twisted bottle}    
\author{Alexandre Arbey \\
Centre de Recherche Astrophysique de Lyon, Observatoire de Lyon, Universit\'e Lyon 1,\\
9 avenue Charles Andr\'e, Saint-Genis Laval cedex, F-69561, France; CNRS, UMR 5574; \\
Ecole Normale Sup\'erieure de Lyon, Lyon, France.\\
CERN, Theory Division, Physics Department, CH-1211 Geneva 23, Switzerland.\\
E-mail: \email{alexandre.arbey@ens-lyon.fr} }
\author{Giacomo Cacciapaglia, Aldo Deandrea, Bogna Kubik \\
   Universit\'e de Lyon, France; Universit\'e Lyon 1,\\ 
   CNRS/IN2P3, UMR5822 IPNL, F-69622 Villeurbanne Cedex, France.\\
E-mails: \email{g.cacciapaglia@ipnl.in2p3.fr, deandrea@ipnl.in2p3.fr, bkubik@ipnl.in2p3.fr}}

\abstract{The real projective plane is a compact, non-orientable orbifold of Euler characteristic 1 without boundaries, which can be 
described as a twisted Klein bottle. We shortly review the motivations for choosing such a geometry among all possible 
two-dimensional orbifolds, while the main part of the study will be devoted to dark matter study and limits in Universal Extra 
Dimensional (UED) models based on this peculiar geometry. In the following we consider such a UED construction based on the direct 
product of the real projective plane with the standard four-dimensional Minkowski space-time and discuss its relevance as a model of a 
weakly interacting Dark Matter candidate. \\
One important difference with other typical UED models is the origin of the symmetry leading to the stability of the dark matter particle. 
This symmetry in our case is a remnant of the six-dimensional Minkowski space-time symmetry partially broken by the 
compactification. Another important difference is the very small mass splitting between the particles of a given Kaluza-Klein tier, which 
gives a very important role to co-annihilation effects.
Finally the role of higher Kaluza-Klein tiers is also important and is discussed together with a detailed numerical description of the 
influence of the resonances.}
\preprint{LYCEN 2012-03, CERN-PH-TH/2012-211} 

\begin{document}

\section{Introduction}
The motivations for building a specific model based on a compactified Extra Dimension can be of different origins, ranging from 
phenomenological ones to more formal ones related for example to string theory. 
For instance, the issue of the radiative stability of the Higgs boson mass or the mass hierarchy in the fermion and gravity sectors can be 
addressed in novel ways compared to 4 dimensional physics.
In this paper we focus on the issue of Dark Matter in Universal Extra Dimension (UED) type of models.
In the following we shall consider an effective theory defined on 
a $d$ dimensional manifold which is the direct product of the standard four-dimensional Minkowski space-time $\mathcal{M}^4$ and a $(d-4)$-dimensional orbifold defined as the  quotient space of $\mathbb{R}^{(d-4)}$ modulo a discrete symmetry group $\Gamma$. This general 
framework can be constrained by few important theoretical requirements which will select, in our case, a unique geometry.  

One of the most attractive explanations to the presence of Dark Matter (DM) in the universe is the existence of a weakly interacting 
particle (WIMP), present in many extensions of the Standard Model.  A main requirement for a particle theory model is therefore the 
possibility to obtain, in a natural way, a viable Dark Matter candidate, in the form of a neutral, stable and weakly interacting particle. 
From a theoretical point of view, its stability should be obtained as the result of a 
symmetry conservation law. In this respect many models impose an {\it ad hoc} parity, which may or may not have other independent 
justifications. 
In extra dimensions, residual symmetries of the compact space can play the role of the parity that stabilises a Dark Matter candidate \cite{Servant:2002aq}, however such symmetry usually requires {\it ad hoc} constraints on the effective Lagrangian of the model.
The scenario we discuss here, based on the real projective plane orbifold, has a special status as the stability of the dark 
matter candidate is not imposed, but is the result of an exact residual space-time symmetry after compactification. 

The presence of exact residual symmetries is related to fixed points or lines. Fixed points, which are points in the extra space which are left invariant by all the symmetries of the orbifold projection, in general break the $d$-dimensional Lorentz invariance to 
the 4-dimensional one, therefore no extra symmetry would survive in general and the predictivity of the model is considerably reduced. Indeed the divergences appearing in loop 
corrections require counter-terms localised on these fixed points. Another important theoretical and phenomenological requirement is 
the presence of 4-dimensional chiral fermions as zero modes in the low energy spectrum of the effective theory. 
The requirements of the absence of fixed points and of the presence of chiral zero modes completely eliminate the possibility of 
working with only one compact extra dimension, as the only 1-dimensional orbifold without fixed points is the circle $S^1$. However no 
chiral fermions can be defined on the circle without taking a quotient introducing fixed points. 
In~\cite{Cacciapaglia:2009pa} it was shown that the unique orbifold without fixed points, among the 17 orbifolds which can be defined 
on a 2-dimensional euclidean plane, is the real projective plane (RP$^2$). One may consider higher dimensional orbifolds, but naive 
dimensional 
analysis shows quite easily that increasing the space-time dimensionality of the effective theory brings automatically an increase in the 
dimension of the fields and operators which reduces drastically the predictivity of an effective theory based on the quantum field theory 
paradigm. We shall therefore limit our study to this unique 2-dimensional orbifold, the real projective plane.
Its fundamental domain is a rectangle with opposite sides identified after being twisted, like a ``double'' M\"obius strip.
If we were to twist only along one direction, and join the other two sides plainly, we would obtain a Klein bottle: therefore, the real projective plane can also be thought as a {\it twisted Klein bottle}.

The paper is organised as follows: after summarising in section \ref{sec:dm} the basic formulas for the DM relic density calculation, in section \ref{sec:model} we briefly present the real projective plane and the corresponding UED 
model. 
 In section \ref{seq:results} we describe in detail the peculiar spectrum of the RP$^2$ model, its dark matter particle candidate (LKP) and the main formulas for its pair annihilation. In the following sections \ref{sec:analytical_results}
and \ref{sec:relicAbundanceBounds} we study the relic abundance using respectively analytic and numerical calculations. These two
sections complete nicely each other for the understanding of the relic abundance behaviour of different contributions: annihilation 
versus co-annihilation and different particles contributions for DM observables. The effect of the cut--off of the effective theory is discussed in section \ref{sec:cutoff}, while that of the localised 
Higgs mass in section \ref{sec:locHiggs}. In section \ref{sec:direct} we consider the present and future direct detection bounds. Section \ref{sec:conclusion} contains our 
conclusions.

\section{Dark Matter relic density}
\label{sec:dm}
In order to compute the dark matter relic density, we assume the cosmological standard model, which is based on a 
Friedmann-Lema{\^\i}tre Universe filled with radiation, baryonic matter and cold dark matter, approximately flat and incorporating a 
cosmological constant accelerating its expansion. Before recombination, the Universe expansion is dominated by a radiation density, 
and therefore the expansion rate $H$ of the Universe is determined by the Friedmann equation
\begin{equation}
H^2=\frac{8 \pi G}{3} \rho_{rad}\;,\label{friedmann_stand}
\end{equation}
where
\begin{equation}
\rho_{rad}(T)=g_{\mbox{eff}}(T) \frac{\pi^2}{30} T^4
\end{equation}
is the radiation density and $g_{\mbox{eff}}$ is the effective number of degrees of freedom of radiation. The computation of the relic 
density is based on the solution of the Boltzmann evolution equation \cite{relic_calculation1,relic_calculation2}
\begin{equation}
dn/dt=-3Hn-\langle \sigma_{\mbox{eff}} v\rangle (n^2 - n_{\mbox{eq}}^2)\;, \label{evol_eq}
\end{equation}
where $n$ is the number density of all KK particles, $n_{\mbox{eq}}$ their equilibrium density, and 
$\langle \sigma_{\mbox{eff}} v\rangle$ is the thermal average of the annihilation rate of the KK particles to the 
Standard Model particles.
The thermal average of the effective cross section is given by
\begin{equation}
\langle \sigma_{\rm{eff}}v \rangle = \dfrac{\displaystyle\int_0^\infty dp_{\rm{eff}} p_{\rm{eff}}^2 W_{\rm{eff}}(\sqrt{s}) K_1 
\left(\dfrac{\sqrt{s}}{T} \right) } { m_{LKP}^4 T \left[ \displaystyle\sum_i \dfrac{g_i}{g_{LKP}} \dfrac{m_i^2}{m_1^2} K_2 \left(\dfrac{m_i}{T}
\right) \right]^2}\;,
\label{eq:sigv_eff}
\end{equation}
where $K_1$ and $K_2$ are the modified Bessel functions of the second kind of order 1 and 2 respectively, and
\begin{equation}
\frac{d W_{\rm eff}}{d \cos\theta} = \sum_{ijkl} \frac{p_{ij} p_{kl}}{ 8 \pi g_{LKP}^2 p_{\rm eff} S_{kl} \sqrt{s} }
\sum_{\rm helicities} \left| \sum_{\rm diagrams}  \mathcal{M}(\tilde{i}\tilde{j} \to kl) \right|^2 \;,
\end{equation}
where $\mathcal{M}(\tilde{i}\tilde{j} \to kl)$ is the transition amplitude of the  (co-)annihilation of KK particles $\tilde{i}$ and $\tilde{j}$ 
into SM particles $k$ and $l$,
\begin{equation}
g_{LKP}^2 p_{\rm{eff}} W_{\rm{eff}} \equiv \sum_{ij} g_i g_j p_{ij} W_{ij}
\end{equation}
with
\begin{equation}
p_{\rm{eff}}(\sqrt{s}) = \frac{1}{2} \sqrt{(\sqrt{s})^2 -4 m_{LKP}^2} \;,
\end{equation}
and where $\theta$ is the angle between particles $\tilde{i}$ and $k$.
By solving the Boltzmann equation, the density number of KK particles in the present Universe and consequently the relic density can 
be determined.
The ratio of the number density to the radiation entropy density, $Y(T)=n(T)/s(T)$ can be defined, where
\begin{equation}
s(T)=h_{\mbox{eff}}(T) \frac{2 \pi^2}{45} T^3 \;.
\end{equation}
$h_{\mbox{eff}}$ is the effective number of entropic degrees of freedom of radiation. Combining Eqs. (\ref{friedmann_stand}) and 
(\ref{evol_eq}) and defining $x=m_{\mbox{\small LKP}}/T$, the ratio of the LKP mass over temperature, yield
\begin{equation}
\frac{dY}{dx}=-\sqrt{\frac{\pi}{45 G}}\frac{g_*^{1/2} m_{\mbox{\small LKP}}}{x^2} \langle \sigma_{\mbox{eff}} v\rangle (Y^2 - 
Y^2_{\mbox{eq}}) \;, \label{main}
\end{equation}
with
\begin{equation}
g_*^{1/2}=\frac{h_{\mbox{eff}}}{\sqrt{g_{\mbox{eff}}}}\left(1+\frac{T}{3 h_{\mbox{eff}}}\frac{dh_{\mbox{eff}}}{dT}\right) \;.
\end{equation}
The freeze-out temperature $T_f$ is the temperature at which the LKP leaves the initial thermal equilibrium when $Y (T_f) = (1 + \delta) 
Y_{\mbox{eq}}(T_f)$, with $\delta \simeq 1.5$. The relic density is obtained by integrating Eq. (\ref{main}) from $x=0$ to 
$m_{\mbox{\small LKP}}/T_0$, where $T_0=2.726$ K is the temperature of the Universe today
\cite{relic_calculation1,relic_calculation2}:
\begin{equation}
\Omega_{\mbox{\small LKP}} h^2 = \frac{m_{\mbox{\small LKP}} s(T_0) Y(T_0) h^2}{\rho_c^0} \approx 2.755\times 10^8 
\frac{m_{\mbox{\small LKP}}}{1 \mbox{ GeV}} Y(T_0)\;,
\end{equation}
where $\rho_c^0$ is the critical density of the Universe, such as
\begin{equation}
H^2_0 = \frac{8 \pi G}{3} \rho_c^0 \;,
\end{equation}
$H_0$ being the Hubble constant. The obtained relic density $\Omega_{\mbox{\small LKP}} h^2$ can then be directly compared to the observed dark matter density. The numerical calculation of the relic density is performed using MicrOMEGAs v2.4.1 \cite{Belanger:2001fz,Belanger:2004yn}.
To constrain the relic density, we consider the 7-year WMAP data (WMAP7), which have provided an unprecedented measurement of the cold dark matter density \cite{Komatsu:2010fb}:
\begin{equation}
 \Omega_{cdm} h^2 = 0.1123 \pm 0.0035 \;.
\end{equation}
Taking into consideration 10\% of theoretical uncertainty in the relic density calculation, we impose the following constraint:
\begin{equation}
0.0773 < \Omega h^2 < 0.1473 \;.
\end{equation}
It is important to remark however that the calculation of the relic density rely on many cosmological assumptions. In particular, different cosmological scenarios can lead to a relic density which is larger than that computed in the standard cosmological scenario. First, the LKP could be only one of several dark matter components. Then, if dark energy were the dominant component at the time of the relic freeze-out, it  would result in an acceleration of the expansion of the Universe, which would lead to an earlier freeze-out and a much larger relic density~\cite{Kamionkowski:1990ni,Salati:2002md,Profumo:2003hq,Chung:2007cn,Arbey:2008kv}. Finally, entropy generation at the time of freeze-out, for example due to the decay of a late inflaton, can also lead to an increase -- or a decrease -- of the relic density~\cite{Moroi:1999zb,Giudice:2000ex,Fornengo:2002db,Gelmini:2006pq,Arbey:2009gt}. These effects are however limited by Big-Bang nucleosynthesis constraints, but using SuperIso Relic \cite{Arbey:2009gu,Arbey:2011zz} and AlterBBN \cite{Arbey:2011nf}, it can be verified that they can nevertheless lead to an increase of three orders of magnitudes or more of the relic density while still being compatible with BBN constraints. For this reason, the lower dark matter density bound can be considered as a weak constraint.

\section{UED on a twisted bottle}
\label{sec:model}

The real projective plane $RP^2$ is a compact, non-orientable orbifold of Euler characteristic 1 without boundaries. 
It can be constructed in two ways, either starting from a sphere $S^2$ or from an infinite plane $\mathbb{R}^2$. The two constructions are not equivalent as in the first case the curvature is distributed on the surface while in the second we have a flat metric $g^{MN} = \mathrm{diag}(1,-1,-1,-1,-1,-1)$ except for two conical singularities (but not fixed points) where curvature is concentrated. The Kaluza-Klein spectrum is also different: starting on the sphere $S^2$, the modes are labelled by angular momentum, while starting on the plane $\mathbb{R}^2$ the modes are labelled by quantised momentum along the two directions.
In this paper we select the simplest case from the point of view of discussing fermions, so we choose the flat version of the orbifold that was described in Ref. \cite{Cacciapaglia:2009pa}. Some aspects of the spherical RP$^2$ are discussed in Ref. \cite{Dohi:2010vc}.

\begin{figure}[htb]
\begin{center}
\includegraphics[scale=0.7]{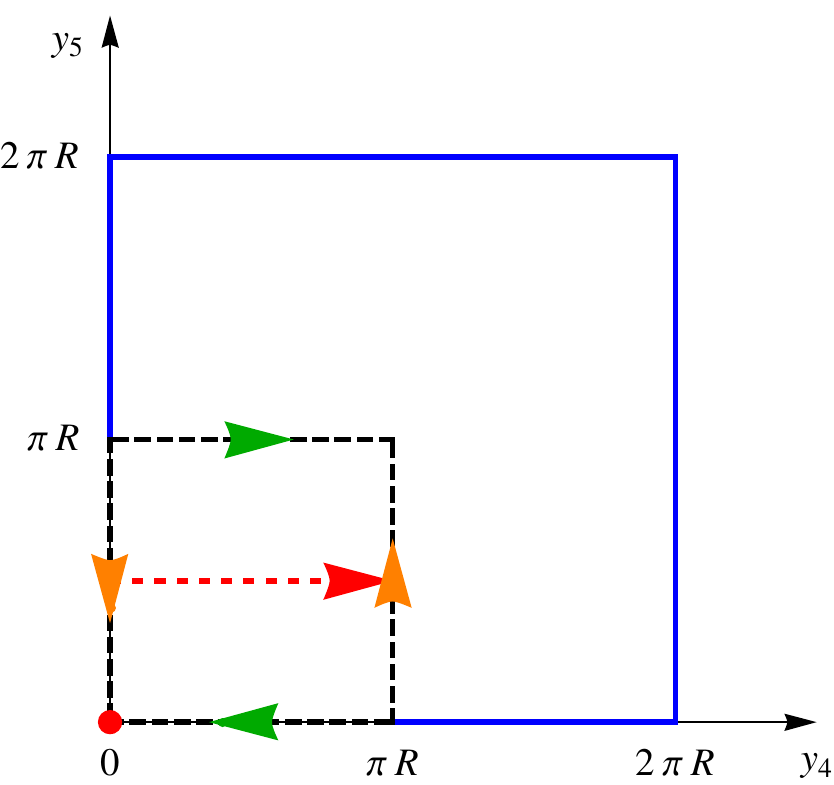}
\includegraphics[scale=0.7]{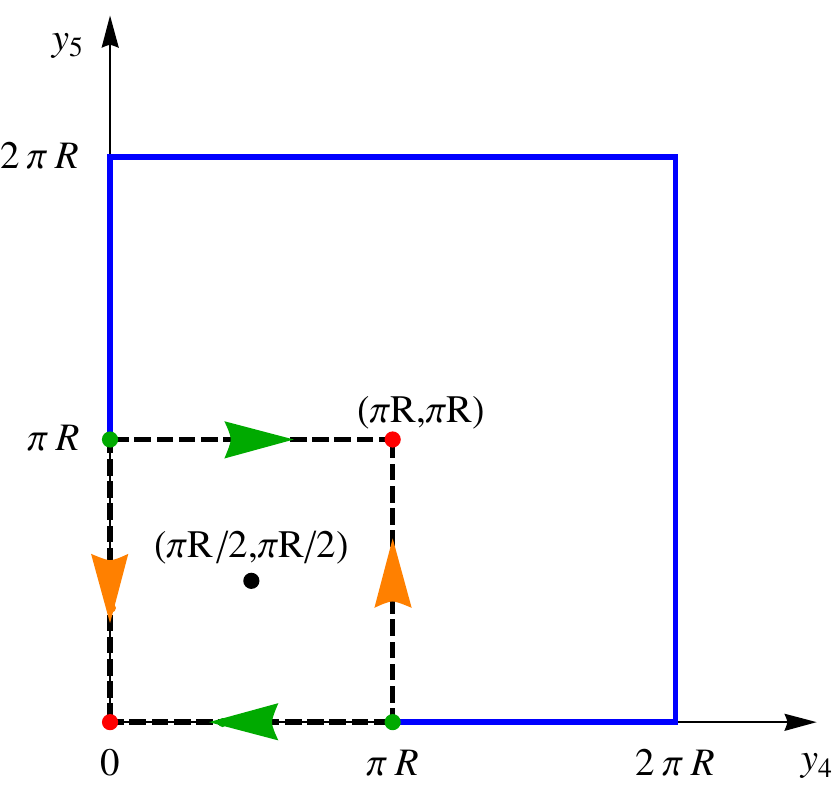}
\end{center}
\caption{\footnotesize On the left panel: Fundamental domain of the real projective plane (within black dashed lines) embedded in a torus (blue square). The red dot is $\pi$-rotation generator; the red dashed arrow represents the glide generator $g$. The green and orange arrows indicate the identification of the facing sides.
On the right panel: Geometrical properties of the real projective plane. The identified pairs of singular points are marked by red and green dots. The centre of the square (black dot) is the $p_{KK}$ parity centre. In this figure we set $R_4 = R_5 = R$.} 
\label{fig:orbifold}
 \end{figure}

The ``flat'' real projective plane is defined as a quotient space RP$^2=\mathbb{R}^2/\Gamma_{RP^2}$ where $\Gamma_{RP^2}$ is a discrete symmetry group defined by two symmetry generators $g$ and $r$ as:
\begin{equation}
 \Gamma_{RP^2} = \langle r,g | r^2 = (g^2r)^2 = 1 \rangle\,.
\end{equation}
We choose a particular representation of the generators in terms of the isometries of the plane\footnote{Note that the structure of the group is entirely defined by the relations between the generators. Their particular representation in terms of the isometries of the plane is not necessary but helps in visualisation.}:

\begin{equation}
\begin{array}{ll}
g:\left\{\begin{array}{l}
x_4 \sim g(x_4) = -x_4 +\pi R_4\\
x_5 \sim g(x_5) =  x_5 + \pi R_5
\end{array}\right.\,, \quad
r:\left\{\begin{array}{l}
x_4 \sim r(x_4) = -x_4 \\
x_5 \sim r(x_5) = -x_5
\end{array}\right. \,,
\end{array}
\end{equation}
so that $g$ generates the glide (mirror reflection + translation) and $r$ corresponds to the rotation of $\pi$ degrees around the origin.
The fundamental domain of our twisted bottle can be visualised as a rectangle of sides of length $\pi R_4$ and $\pi R_5$, 
with opposite sides identified with a twist. 
The orbifold obtained in this way has no fixed points and no fixed lines as can be seen in Figure \ref{fig:orbifold}. 
In fact, the glide makes sure that no point can be identified with itself. The corners of the rectangle, which are fixed points of the rotation, are identified by the glide: $(0,0) \sim (\pi R_4, \pi R_5)$ and $(0,\pi R_5) \sim (\pi R_4,0)$. These 
two physically nonequivalent points form two conical singularities with deficit angle $\pi$, which concentrate the curvature of the real projective plane but keeping the metric finite on these points. 

The two symmetries $g$ and $r$ generate two translations $t_4 \equiv (g * r)^2$ and $t_5\equiv g^2$ along the two directions $x_4$ and $x_5$ respectively under which the space is periodic, therefore the real projective plane can be embedded in a torus $T^2$. This is an important property as the torus is an orientable orbifold and fermions can therefore be defined in the same way as on the $T^2$ orbifold.
The chirality of the 4-dimensional fermions can be properly defined on this space-time geometry thanks to the rotation projection $r$ (see Ref. \cite{Cacciapaglia:2009pa} for details). 

\subsection{Field content}

We are interested in a Universal Extra Dimensions model where all the Standard Model particles are allowed to propagate in the two extra dimensions. We consider a minimal version of the model that can be defined on $RP^2$, that is we introduce only the fields whose zero modes will reproduce the Standard Model content. The detailed construction of the orbifold and quantum fields decomposition is presented in Ref. \cite{Cacciapaglia:2009pa}. Here we mention the important points about the spectrum for the purpose of this paper.

Each quantum field is now a 6-dimensional field which, while projected on the orbifold, decomposes in an infinite tower of massive Kaluza-Klein (KK) modes.
Each tier of modes is labelled by two integers $(k,\,l)$ which correspond to the discretised momenta along the extra directions. 
While projected on the orbifold, the fields can have in general four different parities under orbifold projection: $(p_r,\,p_g) = (\pm 1,\,\pm 1)$. The parities are chosen in such a way that the zero mode spectrum corresponds to the SM.

To each SM field there corresponds exactly one six-dimensional field and therefore one tower of massive KK resonances. The only exception are fermions: we have to introduce two six-dimensional spinors $\Psi^{6D} = (\chi_+,\,\bar{\eta}_-,\,\chi_-,\,\bar{\eta}_+)^T$ with opposite rotation parities for each four-dimensional Dirac spinor. In this notation $\pm$ subscripts correspond to the 6D chiralities while $\chi$, $\eta$ are the 4D chirality eigenstates. The rotation projection will cancel the the zero-more wave functions of left or right handed components ($\chi$ or $\eta$) in $\Psi^{6D}$ assuring a chiral 4D zero mode. For example, for a 4-dimensional Dirac spinor $e^{4D}$ we should introduce a 6-dimensional spinor $e_L^{6D}$ corresponding to a left handed component of $e^{4D}$ and a $e_R^{6D}$ corresponding to a right handed component of $e^{4D}$, and as a result for a SM electron we will have two infinite towers of massive KK states which will differ by rotation parity assignment $p_r$.

The mass eigenstates can be labelled by their parity assignment $(p_r,p_g)$ and KK number $(k,l)$.
As an illustration, the classification of the modes with normalised wave functions for a scalar field from \cite{Cacciapaglia:2009pa} is (here for simplicity we fix $R_4 = R_5 = 1$) :
{\footnotesize
\begin{center}
\begin{tabular}{c|c|c|c|c|c|c|}
 $(k,l)$ & $p_{KK}$  & $(++)$ & $(+-)$ & $(-+)$ & $(--)$ \\
 \hline
 $(0,0)$ & $+$ & $\frac{1}{2 \pi}$ & & & \\
 \hline
 $(0, 2l)$ & $+$ & $\frac{1}{\sqrt{2} \pi} \cos 2 l x_6$ & & & $\frac{1}{\sqrt{2} \pi} \sin 2 l x_6$ \\
 \hline
 $(0,2l-1)$ & $-$  & &  $\frac{1}{\sqrt{2} \pi} \cos (2 l-1) x_6$ & $\frac{1}{\sqrt{2} \pi} \sin (2 l-1) x_6$ & \\
 \hline
 $(2k,0)$ & $+$  & $\frac{1}{\sqrt{2} \pi} \cos 2 k x_5$ &  & $\frac{1}{\sqrt{2} \pi} \sin 2 k x_5$ & \\
\hline
 $(2k-1,0)$ & $-$ & & $\frac{1}{\sqrt{2} \pi} \cos (2 k-1) x_5$ & & $\frac{1}{\sqrt{2} \pi} \sin (2 k-1) x_5$ \\
\hline
$(k,l)_{\rm k+l\; even}$ & $+$  & $\frac{1}{\pi} \cos k x_5 \cos l x_6$ & $\frac{1}{\pi} \sin k x_5 \sin l x_6$ & $\frac{1}{\pi} 
\sin k x_5 \cos l x_6$ & $\frac{1}{\pi} \cos k x_5 \sin l x_6$ \\
\hline
$(k,l)_{\rm k+l\; odd}$ & $-$ & $\frac{1}{\pi} \sin k x_5 \sin l x_6$ & $\frac{1}{\pi} \cos k x_5 \cos l x_6$ & $\frac{1}{\pi} \cos k x_5 \sin l x_6$ & $\frac{1}{\pi} \sin k x_5 \cos l x_6$ \\
\hline
\end{tabular}
\end{center}}

The fundamental space in Figure~\ref{fig:orbifold}, including boundaries and corners, is invariant under a $\pi$-rotation around the centre of the rectangle $r'$. Under this rotation, however, the opposite chiralities will pick a different parity under rotation $r'$ therefore it can not be a good KK parity. An equivalent symmetry can however be defined in terms of translation (see Ref. \cite{RP2spectrum} for details):

\begin{equation}
\begin{array}{ll}
p_{KK} = r'*r:\left\{\begin{array}{l}
x_4 \sim p_{KK}(x_4) = x_4 + \pi R_4\\
x_5 \sim p_{KK}(x_5) = x_5 + \pi R_5
\end{array}\right.\,.
\end{array}
\end{equation}

This exact symmetry of the space can be translated into a parity on the KK states as all the fields in a $(k,l)$ mode will pick up the same phase $(-1)^{k+l}$, therefore modes with even $k+l$ are even and modes with odd $k+l$ are odd. This parity ensures the stability of the lightest odd states as they can never decay into a pair of lighter states which are even under KK parity.
It follows that the lightest odd state belongs to either $(1,0)$ or $(0,1)$, thus it is in such tiers that we will look for a suitable Dark Matter candidate.

\section{Particle Spectrum and the Dark Matter Candidate}
\label{seq:results}

In this section we present the detailed spectrum of the $(1,0)$-$(0,1)$ and $(2,0)$-$(0,2)$ tiers at one loop level. We see that the first KK excitation of the photon $A^{(1,0)}$ and/or $A^{(0,1)}$ is the viable Dark Matter candidate. We emphasise also an interesting property of the spectrum in our model, that all the particles in the same tier are nearly degenerated in mass even after the introduction of radiative corrections. This feature is important for the calculation of the relic abundance of the LKP, since the abundance is strongly affected by co-annihilation processes. 
Note that in other UED models \cite{Dobrescu:2004zi,Burdman:2005sr} the mass splitting are naturally larger.

At leading order,  all the states in each tier are degenerate with mass determined by the two integers $(l, k)$ as
\begin{equation}
m^2_{l,k} = \frac{l^2}{R_4^2} + \frac{k^2}{R_5^2}.
\end{equation}

Splittings within the modes in each tier (k,l) can be generated by three mechanisms: the Higgs vacuum expectation value (VEV), 
bulk interaction loop corrections and higher order operators localised on the singular points.

At one loop order the mass of an $(n,0)$ or $(0,n)$ state can be generically written as
\begin{equation}
m^2_{(n,0)} = \frac{1}{R_{4/5}^2} \left( n^2 + m_{SM}^2 R_{4/5}^2 + \delta_{\rm finite}^{(n,0)} (R_4, R_5) + n^2 \delta_{log} \right)\,, \label{eq:loopmass}
\end{equation}
where $m_{SM}$ is the mass of the correspondent SM state, $\delta_{\rm finite}$ is a finite contribution while $\delta_{log}$ is a divergent contribution depending on the log of a cut-off scale $\delta_{log} \sim \log \Lambda R_{4/5}$. Detailed formulas for the mass corrections for the tiers $(1,0)$ and $(0,1)$ can be found in Ref. \cite{Cacciapaglia:2011hx}, while formulas for the even tiers $(2,0)$ and $(0,2)$ can be found in Ref. \cite{RP2spectrum}. We see that at loop level the masses depend on 3 free parameters: the two radii $R_4$ and $R_5$, and the cut-off $\Lambda R$.
The only exceptions are the massive Higgs modes, which are only present in the even tier $(2,0)$ and $(0,2)$: for the scalar field the most generic Lagrangian contains a mass term $\sim m_{loc}$ localised on the singular points therefore the (2,0) KK scalar mass at loop level is given by \cite{RP2spectrum}:

\begin{equation}
m^2_{H^{(n,0)}} = \frac{1}{R_{4/5}^2} \left( n^2 + m_{SM}^2 R_{4/5}^2 + m_{loc}^2 R_{4/5}^2 + \delta_{\rm finite}^{(n,0)} (R_4, R_5) + 4 \delta_{log} \right)\,, \label{eq:loopmassH}
\end{equation}
where $m_{loc}$ is a free parameter of the model.

As it can be seen in the formulas \ref{eq:loopmass} and \ref{eq:loopmassH} the finite loop corrections $\delta_{\rm finite}^{(n,0)}$ depend on whether we consider the $(n,0)$ or $(0,n)$ mode. However the mass correction is dominated by the log-divergent term, which is enhanced for heavier tiers while the $\delta_{\rm finite}^{(n,0)}$ dependence on $R_4/R_5$ is very mild\cite{RP2spectrum}, thus here for simplicity we consider that $\delta_{\rm finite}^{(n,0)} = \delta_{\rm finite}^{(0,n)}$ and the correction relative to the tree level one is almost independent on the value of the radii $R_4$ and $R_5$.

Moreover we should consider a generic cross level mixing between the modes with different KK numbers: $(n,0)-(m,0)$, $(0,n)-(0,m)$ and $(n,0)-(0,m)$. Mixing for $n\neq m$ can be safely neglected. For $n$ odd the modes $(n,0)-(0,n)$ cannot mix via loops as the vertexes would break the KK parity, therefore only the even modes $(n,0)-(0,n)$ can mix and the mixing angle depends on the ratio $R_4/R_5$. In our numerical study we will focus on two limiting cases: the degenerate case $R_4 = R_5$ and the decoupling limit $R_4 \gg R_5$ and we describe the properties of the spectrum in each of the cases below.

\subsection{Decoupling limit $R_4 \gg R_5$}

In the decoupling limit $R_4 \gg R_5$, all the states $(0,n)$ and $(n,m)$, whose tree level masses have a contribution proportional to $1/R_5^2$ decouple from the spectrum. Therefore, for our purposes, only $(n,0)$ states will contribute. Moreover, the spectrum of the tiers is only mildly dependent on the value of $R_5$ via the finite contributions $\delta_{\rm finite}^{(n,0)}$, therefore as a good approximation we can use the formulas for the mass corrections with $R_4 = R_5$. In this limit no mixing between even modes appears.
Regarding the relic abundance calculation, in fact, the potential Dark Matter candidate from $(0,1)$ is irrelevant as long as the difference in mass $1/R_5 - 1/R_4$ is or order a few times the freeze-out temperature.
As the typical freeze-out temperature is of order few tens of GeV, the decoupling limit is reached even for mildly asymmetrical radii.
Therefore in the decoupling limit the model will contain only one even tier, labelled by a superscript ``(2)'', and one odd tier, labelled by a superscript ``(1)''. 

\subsection{Degenerate radii $R_4 = R_5$.}

In the symmetric $R_4 = R_5 = R$ case, the masses of the two tiers $(0,1)$ and $(1,0)$ are exactly degenerated: in this case we have two dark 
matter particles, $A^{(0,1)}$ and $A^{(1,0)}$, with exactly the same masses and spins. 
States in the two degenerate tires cannot annihilate each other, therefore, in the relic abundance calculation, they must be treated as independent. 
They can scatter via states $(1,1)$, however such processes will only contribute to the thermalisation of the Dark Matter states.
This situation is true up to the presence of localised operators: in fact, adding different operators on the two singular points breaks the degeneracy and generates both mass mixing between $(1,0)$ and $(0,1)$ states and the possibility of the direct coupling of $(1,1)$ states to SM ones.
In the following however we will work in the approximation where such localised operators are absent or negligible.
In the relic abundance calculation we will consider a single odd tier, and multiply the final results by 2.

The situation is different for the odd tiers: in fact, loop corrections can generate mass mixing between $(2,0)$ and $(0,2)$ states. Now the diagonal mass corrections are equal $\delta_{\rm finite}^{(n,0)} = \delta_{\rm finite}^{(0,n)} = \delta_{\rm finite}^{(n)}$ and the off diagonal corrections $\delta'$ introduce a mixing between the two degenerate states. The off diagonal terms can be calculated by use of the localised counter-terms~\cite{Cacciapaglia:2011hx} and they are equal to the log-divergent diagonal term \cite{RP2spectrum}.
As a consequence, the mass eigenstates are given by $A_{(2\pm)}\propto A^{(2,0)} \pm A^{(0,2)}$ with the mass eigenstates equal to
\begin{equation}
m^2_{n\pm)} = \frac{1}{R^2} \left( n^2 + m_{SM}^2 R^2 + \delta_{\rm finite}^{(n)}(R) + n^2 \delta_{log} \pm \delta'\right)\,, \label{eq:loopmassmixing}
\end{equation}
with $\delta'=n^2\delta_{log}$.
Therefore, the sum and difference eigenstates correspond to eigenvalues with by double or no log-divergent terms.
In other words, only one of the mass eigenstates, $A^{(2+)}$, will have log-divergent contribution to the masses while the $A^{(2-)}$ will have no log-divergent contribution. 
Moreover, only the $A^{(2+)}$ plays a significant role in the relic abundance calculation, as it is the only one to couple to a pair of SM states via divergent loop contributions.
In our model implementation, therefore, we only include this tier \cite{RP2spectrum}. 
Therefore we can study the degenerate case by considering one odd tier and one even tier with doubled log-divergent contribution to the mass, and finally multiply by 2 the result of the relic abundance. As in the asymmetric limit the two tiers will be labelled for simplicity with the superscripts ``(1)'' and ``(2)'' respectively.

\subsection{Full numerical spectra and bound on parameters of the model.}

In the following we will focus on the lightest odd tiers $(1,0)$ and $(0,1)$ and the next even tiers $(2,0)$ and $(0,2)$\footnote{The even tier $(1,1)$ may also be relevant, however it only enters in the elastic scattering of $(1,0)$ states off $(0,1)$, therefore it can be safely ignored in this discussion.}. 
As shown in the previous section the mass spectrum at loop level depends on four parameters: the two radii of extra dimensions $R_4$ and $R_5$ as well as on the cut-off scale $\Lambda$ and, for the scalar field only, on the localised mass parameter $m_{loc}$. 

\begin{figure}[!ht]
\begin{center}
\includegraphics[scale=1]{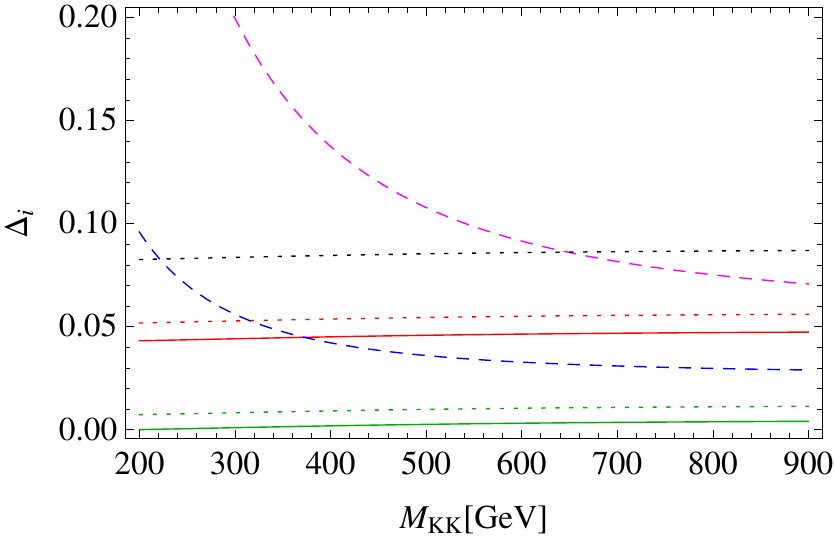}
\caption{\footnotesize First KK level mass splittings as a function of $m_{KK}$ relative to the lightest state ($A^{(1)}$). From bottom to the top: right handed leptons (green), left handed leptons (green dotted), electroweak gauge bosons (blue dashed), singlet light quarks (red), doublet light quarks (red dotted), tops (magenta dashed), gluons (black dotted). }
\label{fig:Mass_Splittings_variations_with_mkk_200_1000}
\end{center}
\end{figure}

\begin{figure}[!ht]
\begin{center}
\includegraphics[scale=0.7]{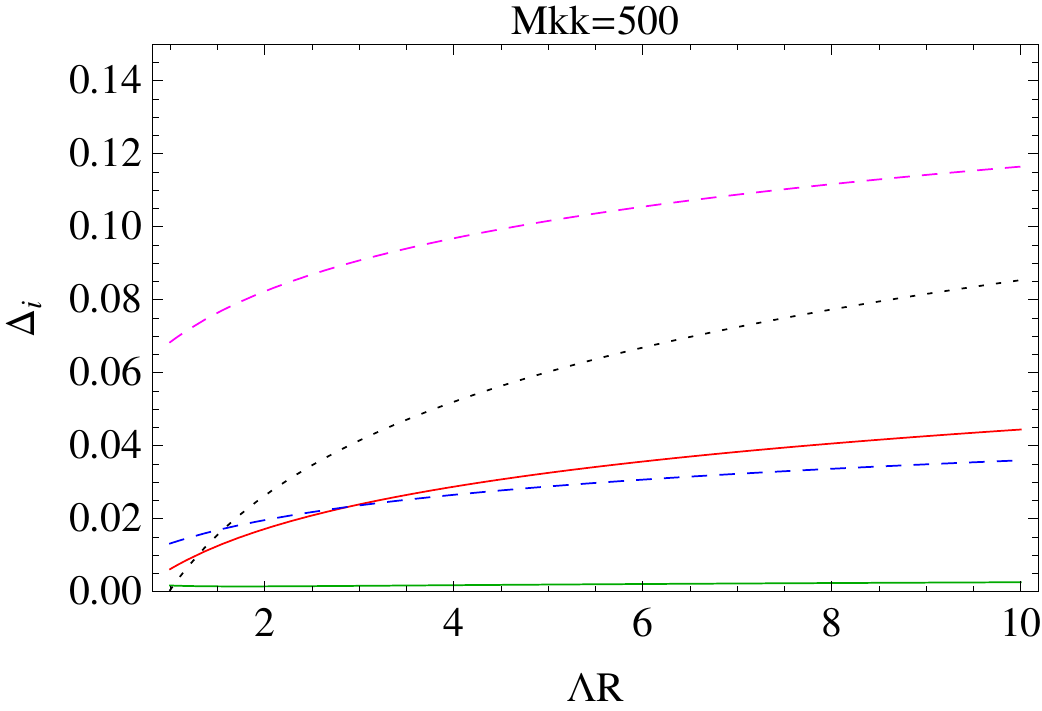}
\includegraphics[scale=0.7]{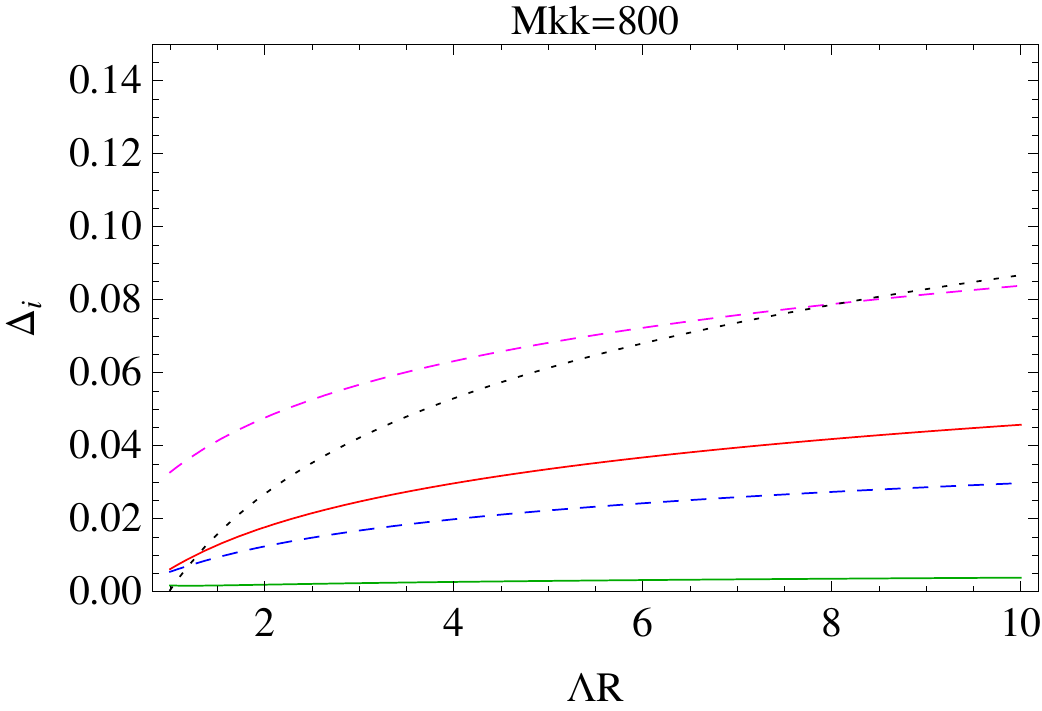}
\caption{\footnotesize First KK level mass splittings as a function of $\Lambda R$. On the left panel: splittings for $m_{KK}=500$ GeV, on the right panel: splittings for $m_{KK}=800$ GeV. On both panels we adopt the colours from bottom to the top at $\Lambda R=10$: leptons (green),  electroweak gauge bosons (blue dashed), light quarks (red), tops (magenta dashed), gluons (black dotted).}
\label{fig:Mass_Splittings_variations_with_LamR}
\end{center}
\end{figure}

The radii $R_4$ and $R_5$ which set the characteristic scale of the spectrum are to be determined by the relic abundance calculation which should be compared with the available cosmological data WMAP $0.0773 < \Omega h^2 < 0.1473$ \cite{Komatsu:2010fb}. 
On the other hand the mass splittings crucially depend on the cut-off $\Lambda$ of the effective 6D model which enters in the logarithmically divergent term $\delta_{log}$. Naive dimensional analysis allows us to estimate the cut off to be a few times $m_{KK}$, up to about a factor of 10.
In Figure \ref{fig:Mass_Splittings_variations_with_mkk_200_1000} we show the relative mass splitting $\Delta_i = \frac{m_i}{m_{A^(1)}}-1$ with respect to the Dark Matter candidate for the odd tier as a function of $m_{KK}$.
The corrections range from $\sim 0.2\%$ for the leptons to a $\sim10\%$ for strongly interacting states, as it is also clear in Table \ref{tab:mass_splittings}, thus justifying the necessity to consider co-annihilation in the relic abundance calculation.
The mass corrections show a mild logarithmic dependence on the cut-off, as shown in Figure \ref{fig:Mass_Splittings_variations_with_LamR}.

\begin{table}[!ht]
\begin{center}
\begin{tabular}{|cc|cccccccc|}
\hline
$m_{KK}$ & $\Lambda R$& $l^{(1)}_R$& $l^{(1)}_L$& $q^{(1)}_S$& $q^{(1)}_D$& $t^{(1)}_S$& $t^{(1)}_D$& $Z^{(1)}$, $W^{(1)}$  & $G^{(1)}$  \\
\hline
 500             & 10         & 0.0026 & 0.0099 & 0.0458 & 0.0545 & 0.1078 & 0.1165 & 0.0373 & 0.0854 \\
 500             & 5          & 0.0020 & 0.0073 & 0.0336 & 0.0399 & 0.0952 & 0.1016 & 0.0288 & 0.0602 \\
\hline
 800            & 10         & 0.0039 & 0.0111 & 0.0471 & 0.0558 & 0.0751 & 0.0838 & 0.0298 & 0.0867 \\
 800            & 5         & 0.0030 & 0.0083 & 0.0346 & 0.0409 & 0.0619 & 0.0682 & 0.0223 & 0.0613 \\
\hline
\end{tabular}
\caption{{\footnotesize Relative mass splittings $\Delta_i = \frac{m_i}{m_{A^1}}-1$ in the L1 model for two values of $m_{KK} = 500,\, 800$ GeV and for two different values of the cut-off $\Lambda R=10$ and $\Lambda R=5$.}}
\label{tab:mass_splittings}
\end{center}
\end{table}

The $m_{loc}$ parameter is another free parameter of the model. It corresponds to the Higgs mass operator localised on the singular points of the orbifold. Note that in principle such mass term might be wrong-signed, i.e. $m_{loc}^2 < 0$, and thus help trigger the electroweak symmetry breaking. This parameter plays a crucial role in the relic abundance calculation, as it enters the Higgs mass (see Eq.~\ref{eq:loopmassH}) and thus can change the position of the $H^{(2)}$ resonance. 
Effective theory order of magnitude estimates suggest that reasonable values for the localised term should be $m_{loc}< m_{KK}$. This estimate is a good starting point as it allows us to expand for small $m_{loc}/m_{KK}$. It is however important to have a more precise limit on $m_{loc}$ than just the effective theory estimate $m_{loc}< m_{KK}$.

\begin{figure}
\begin{center}
 \includegraphics[scale=0.8]{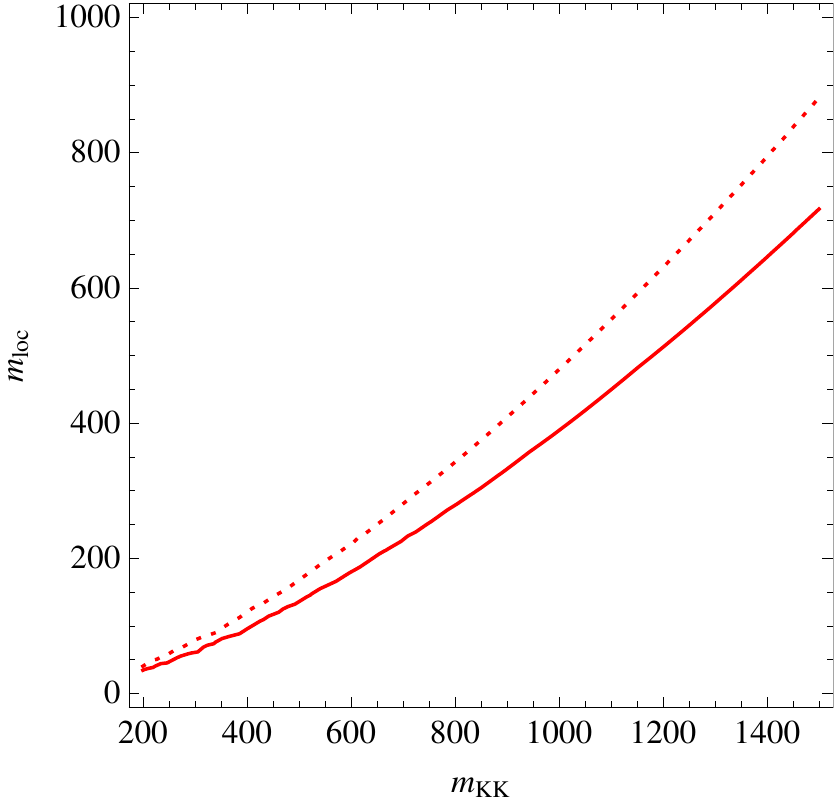}
\caption{Bound on $m_{loc}$ as a function of the $m_{KK}$ mass, obtained imposing that the effect on the $\rho$ parameter is not larger than the allowed measured value within 3 sigmas. Plain line correspond to the bound in the symmetric scenario $R_4 = R_5$, while dashed line to the asymmetric case $R_4 \gg R_5$. }
\label{fig:boundmloc}
\end{center}
\end{figure}

The most relevant bound on $m_{loc}$ comes from electroweak precision measurements, in particular from the $\rho$ parameter.
This can be understood as follows: deriving from localised terms, $m_{loc}$ induces a mass 
mixing of the SM Higgs (zero mode) with all the heavy KK Higgses. In turn, after the SM Higgs boson develops a vacuum expectation value (VEV) $v$, the tadpole generated by the above mixing will propagate the VEV to each massive KK Higgs in the form
\begin{equation}
<H_{(n,0)}> = - \frac{m_{loc}^2}{m_{KK}^2} \frac{v}{n^2}\,, \quad <H_{(0,m)}> = - \frac{m_{loc}^2}{\xi^2 m_{KK}^2} \frac{v}{m^2}\,,   \quad <H_{(n,m)}> = - \frac{m_{loc}^2}{m_{KK}^2} \frac{\sqrt{2} v}{n^2 + \xi^2 m^2}\,; 
\end{equation}
where $m_{KK} = 1/R_4$ is the mass scale related to the larger radius and $\xi = R_4/R_5 \geq 1$.
The bulk kinetic term of the Higgs contains quadrilinear terms between the standard Higgs boson, the higher level Higgs bosons 
and the corresponding $W$'s or $Z$'s. When these Higgses are set to the VEV, we automatically get a mass mixing between the SM $W$ and 
the heavy $W$'s (and similar mixing for the SM $Z$ and the heavy $Z$'s). These effects are not suppressed by the cut-off as we deal with 
$m_{loc}$ which is a dimension 6 operators in 6 dimensions: the only suppressions are due to negative powers of the heavy mass 
scale $m_{KK}$.
These mixings will in turn correct the masses of the $W$ and $Z$, thus potentially affect the $\rho$ parameter.
Other corrections are also generated to the $S$ parameter, however they turn out to be suppressed by extra powers of $m_{KK}$.
The resulting correction to the $\rho$ parameter is
\begin{equation}
\delta \rho = \frac{m_W^2}{m_Z^2 \cos^2 \theta_W} - 1 = -8 \left(\frac{m_{loc}^2}{m_{KK}^2}\right)^2  \frac{m_Z^2-m_W^2}{m_{KK}^2} f(\xi)
\end{equation}
where $\xi = \frac{R_4}{R_5}$ and $f(\xi)$ is a number of order 1, given by the function
\begin{equation}
f (\xi) = \frac{\pi^6}{945} \frac{\xi^6+1}{\xi^6} + 2 \sum_{n,m=1}^{\infty} \frac{1}{(n^2+\xi^2 m^2)^3}\,.
\end{equation}
Numerically, $f (\xi)$ smoothly decreases with increasing $\xi$ and varies from $f(1) = 2.33$ to $f(\infty) = 1.017$.
We finally compare the correction $\delta \rho$ (which is negative in this model) to the value given in PDG $\rho = 1.0004^{+ 0.0003}_{ - 0.0004}$ and obtain the corresponding bounds in the $m_{loc}/m_{KK}$ plane at $3 \sigma$, which are shown in Figure \ref{fig:boundmloc}.
Note that the bound is independent on the sign of $m_{loc}^2$.

The localised counter-terms, that encode the divergent loop corrections to the mass, generate additional corrections to the $\rho$ parameter: for instance log divergent mass corrections to the Higgs will generate a similar contribution to the $\rho$ parameter, while corrections to the Higgs kinetic terms can generate mixing between light and heavy gauge bosons via the SM Higgs VEV.
Such corrections are suppressed by a loop factor compared to $m_{loc}$, however they may be numerically relevant for the extraction of a precise bound on $m_{loc}$.
This task would however require a full one-loop study of the corrections to electroweak precision measurements, which is beyond the scope of this paper.
In this section, we will take the bound on $m_{loc}$ as an indicative value, assuming that there will be no important cancellations with the loop-induced corrections.

\begin{figure}[!ht]
 \begin{center}
    \includegraphics[scale=0.85]{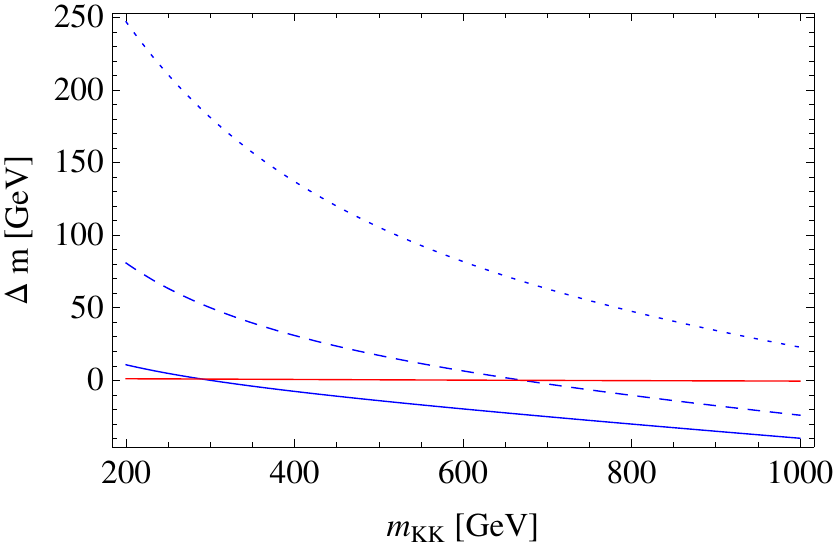}
    \includegraphics[scale=0.85]{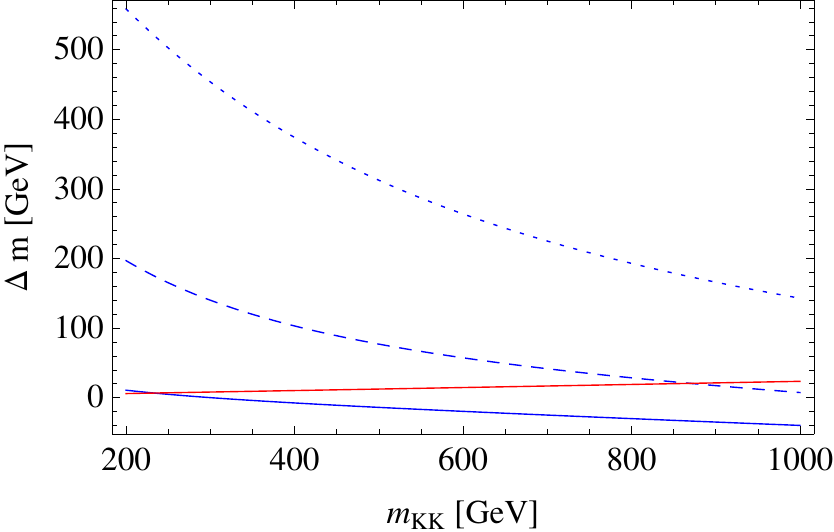}
    \caption{\footnotesize Mass splitting $\Delta m=m - m_{tree}$ in the Higgs sector in tier $(2)$. We show the splittings of $H^{(2)}$ for three values of $m_{loc}$ parameter: $m_{loc}=0$ GeV in blue, $m_{loc}=250$ GeV in blue dashed and $m_{loc}=500$ GeV in blue dotted. On the left non-symmetric case $R_4 \gg R_5$. On the right symmetric case $R_4 = R_5$. Red line corresponds to the $A^{(2)}$ splitting for comparison.}
    \label{fig:Higgs_splittings}
 \end{center}
\end{figure}

\subsubsection*{Full spectrum for the first and second tiers}

In Tables \ref{tab:mass_spectrum_values_10} and \ref{tab:mass_spectrum_values_20} we give values of masses of all the 
particles present in the model for three benchmark points which will be important for the phenomenological analysis: 
$m_{KK} = 300,~500,~ 800$ GeV. The radiative corrections to the KK masses depend on whether we assume $R_4 = R_5$ or 
$R_4 \gg R_5$ only in the second tier $(2,0)-(0,2)$ as there are cross-level mixing terms in this case. We give the mass spectrum in 
both cases.
\begin{table}[!ht]
	\begin{center}
		\begin{tabular}{|c|c|c|c|}
			\hline
			$m_{KK}$ [GeV]& 300 & 500  & 800 \\
			\hline
			$A^{(1)}$ &300.9& 500.5 & 800.2\\
			$l^{(1)}_S$    &301.3& 501.5 & 803.0\\
			$l^{(1)}_D$    &303.4& 504.5 & 808.4\\
			$d^{(1)}_S$    &313.8 & 519.2 & 836.6\\
			$u^{(1)}_S$    &314.2 &519.7 & 837.6\\
			$W^{(1)}$      &317.8& 515.0 & 822.1\\
			$Z^{(1)}$      &319.6 & 515.8 & 822.3\\
			$q^{(1)}_D$    &316.8& 522.9 & 843.9\\
			$t^{(1)}_S$    &361.0 & 550.5 & 859.8\\
			$t^{(1)}_D$    &363.6& 553.6 & 866.1\\
			$G^{(1)}$      &326.1& 534.9 & 868.4\\
			\hline
		\end{tabular}
\caption{\footnotesize{Typical masses of the particles of the level (1) at $m_{KK}=300~,500$ and 800 GeV. The mass splittings are independent on the symmetric or asymmetric case as there is no cross-level mixing.}}
\label{tab:mass_spectrum_values_10}
	\end{center}
\end{table}

\begin{table}[!ht]
	\begin{center}
		\begin{tabular}{|c|c|c|c|}
			\hline
			$m_{KK}$ [GeV]& 300&500 & 800 \\
			\hline
			$S^{(2)}$       &600.0 & 993.4  & 1568.3\\
			$A^{(2)}$     & 600.9& 1000.3 & 1599.8\\
			$H^{(2)}$       &600.0 & 1001.3 & 1573.3\\
			$l^{(2)}_S$     &602.3& 1001.2 & 1605.4\\
			$l^{(2)}_D$     &606.2& 1003.2 & 1615.0\\
			$W^{(2)}$       & 619.6& 1010.5 & 1636.8\\
			$Z^{(2)}$       &620.0& 1010.8 & 1636.8\\
			$d^{(2)}_S$     &625.0& 1014.7 & 1665.5\\
			$u^{(2)}_S$     &625.7& 1015.0 & 1667.3\\
			$q^{(2)}_D$     &630.4& 1017.5 & 1678.6\\
			$t^{(2)}_S$     &651.9 & 1031.3 & 1682.0\\
			$G^{(2)}$       &653.1& 1031.8 & 1739.2\\
			$t^{(2)}_D$     &656.6& 1033.7 & 1693.3\\
			\hline
		\end{tabular}
		\begin{tabular}{|c|c|c|c|}
			\hline
			$m_{KK}$ [GeV] & 300&500 & 800  \\
			\hline
			$S^{(2)}$       &573.3 &  955.6  &1528.9  \\
			$A^{(2)}$     & 601.0& 1000.2& 1599.5\\
			$H^{(2)}$       & 586.8&963.7 & 1534.0\\
			$l^{(2)}_S$     & 604.5& 1007.5 & 1612.0 \\
			$l^{(2)}_D$     &612.3&1020.6 &1632.9 \\
			$W^{(2)}$       & 633.5 & 1050.5  &1677.9 \\
			$Z^{(2)}$       &633.8 &1050.6 & 1677.9 \\
			$d^{(2)}_S$     &659.9 & 1094.4 & 1744.0\\
			$u^{(2)}_S$     &661.4 &1096.9  & 1748.0\\
			$q^{(2)}_D$     & 670.7 &  1112.5 & 1772.9 \\
			$t^{(2)}_S$     &688.4  & 1118.2 &1768.7  \\
			$G^{(2)}$       &  721.1&  1191.8 & 1893.5 \\
			$t^{(2)}_D$     & 697.7 &  1133.8  &  1793.6 \\
			\hline
		\end{tabular}
\caption{\footnotesize{Typical masses of the particles of the level (2) at $m_{KK}=300~,500$ and 800 GeV.  On the left the
non-symmetric case $R_4 \gg R_5$.  On the right the symmetric case $R_4 = R_5$. In both cases, for the Higgses $m_{loc}=0$.}}
\label{tab:mass_spectrum_values_20}
	\end{center}
\end{table}

In the first tier the lightest state is always $A^{(1)}$, which is a neutral spin-0 particle and corresponds to the Dark Matter candidate in our model. For $m_{KK}<200$ GeV the lightest particle in the first tier is the singlet electron which excludes the model in this range of $m_{KK}$.
$A^{(1)}$ is the KK resonance of the photon, however the mixing angle between SU(2) and U(1) components is different from the SM one.
In particular, it is smaller than the Weinberg angle and, for large $m_{KK}$, the DM candidate is predominantly a U(1) gauge boson.
In the degenerate case, two such particles with the same mass exist while in the decoupling limit only the lightest one will significantly contribute to the Dark Matter abundance.
The next lightest 
particles are singlet and doublet leptons and the SU(2) gauge bosons which will play an important role in the dark matter 
phenomenology in the co-annihilation processes.  Coloured particles are heavier but still their role cannot be neglected in the dark matter 
prediction and in collider phenomenology as they have strong couplings which will enhance all the processes where quarks are 
involved. 
Note that there may be additional contributions from the cut-off scale physics to operators localised at the two singular points of the fundamental domain. In principle, these could turn some other particle into the lightest KK-odd state. Hence, the odd modes of neutrinos $\nu^{(1)}$ or the neutral electroweak gauge 
boson $Z^{(1)}$ could all be viable dark matter candidates too. We leave the investigation of these possibilities for future work.

In the second tier, the lightest particle is a vector gauge boson $A^{(2)}$ and Higgs excitations $H^{(2)}$ and $S^{(2)}_{\pm,0}$, if the localised mass parameter $m_{loc}$ 
is very small or $m_{loc}^2 < 0$. The dependence of the Higgs resonance masses to $m_{loc}$ is shown in Figure \ref{fig:Higgs_splittings}. Due to the extremely small mass corrections, $A^{(2)}$ can only decay into SM particles via loop induced interactions and 
thus will play an important role in the Dark Matter and LHC phenomenology enhancing the resonant productions of SM particles. The weak gauge 
bosons $W^{(2)}$ and $Z^{(2)}$ will decay into other heavy particles but the resonant decays into SM fermions will not be negligible as 
well. All the level $(2)$ particles will also participate as final states of the annihilations and co-annihilations of the primordial cosmic plasma 
reducing significantly the relic abundance of the dark matter and thus changing strongly the bounds of the cosmologically allowed 
$m_{KK}$ values \cite{Kakizaki:2005uy}. We will present those phenomenological aspects of the model in the following sections.

Note 
that, in contrast to the chiral square \cite{Dobrescu:2004zi,Burdman:2005sr} and to the 5D mUED \cite{UED5,Kakizaki:2006dz} models where the odd level Higgs boson 
is also a viable dark matter candidate in 
some parameter space, the Higgs boson of level $(1,0)$ and $(0,1)$ mode is not present on the RP$^2$.
In the following, we will neglect the contributions of the localised operators under the assumption that their contribution is smaller that the loop ones.
We will first discuss some features of the two limiting cases under study and the numerical results for the spectra before turning our attention to the calculation of the relic abundance.

\section{Analytical Results}
\label{sec:analytical_results}


In our computation we consider a general particle spectrum without any simplifying assumptions. In particular we do not assume a 
completely degenerate particle spectrum but we keep the KK masses after one-loop corrections to the $(1)$ modes (the analytic formulas, to simplify the notation, will be however shown assuming the 
masses at each KK level degenerate: i.e. for the first level particles $(1)$ we assume $m_{X^{(1)}}=m_{A^{(1)}}$ and for the second level 
particles $(2)$ the mass will be $m_{X^{(2)}}=2 m_{A^{(1)}}$ for every particle $X$ in a given tier).
We keep all the SM particle masses non zero, except the electron mass which will be neglected. Moreover we do not neglect the mixing 
between the U(1) $B^{(1)}$ and SU(2) $W^{(1)}_3$ components, which originates from the electroweak symmetry breaking and is expected to be small for large 
$m_{KK}$ but effectively is important for small values of $m_{KK}$.

In our analytic and numerical calculations we make some simplifying assumptions for the Yukawa couplings. We neglect all the 
Yukawa couplings, which are proportional to the corresponding fermion mass, for all the light SM particles except the top quark. This 
assumption is well justified as the top Yukawa coupling can alter significantly the cross section result as it leads to resonant s-channel 
diagrams. Those resonant effects are important only for the exchange of the $(2)$ level particles and as we will see, they enhance 
the effective cross section in a small region of $m_{KK}$ near the resonant value. Aside those effects, we do not expect our results to be 
sensitive to the masses of light fermions.

In all the analytic calculations when examining the relic density as a function of mass $m_{KK}$ we neglect the mass dependence of 
$x_F$. The value of $x_F$ depends weakly on the mass. This dependence is approximately logarithmic. Typically, over the mass 
range $m_{KK}=0.2-1$ TeV, $x_F$ varies by about 0.1 GeV/degree, or less than 15\%. This variation has small effects on the relic 
density. This also shows that the dark matter is cold. In all cases considered here we obtain $22 \lessapprox x_F \lessapprox 30$ GeV, 
so that the particles are well approximated as non-relativistic. This implies freeze-out 
temperatures in the range $34\div 45$ GeV depending on the scenario.

\subsection{LKP annihilations - analytic expressions for the cross sections}

In this section we show the calculation of annihilation cross section in details in order to fix the notation. The dark matter candidate can annihilate into all SM particles. As we will see the cross sections into SM gauge bosons give the leading contributions. The fermionic 
final states will not contribute once we develop the cross section for small velocities.

\subsubsection{$A^{(1)}A^{(1)}\rightarrow ZZ$}
\begin{figure}[!ht]
\begin{center}
     \includegraphics[scale=0.4]{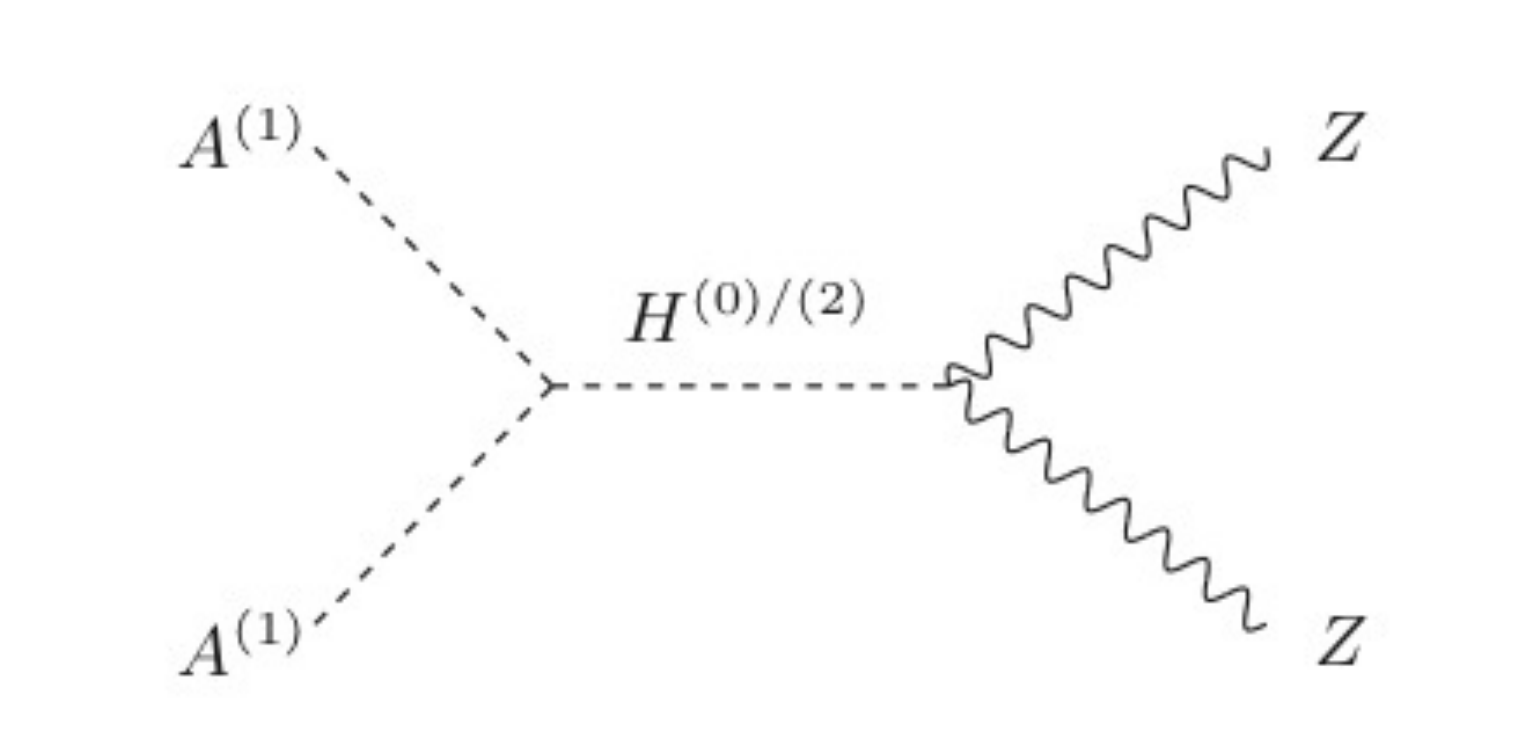}
\caption{\footnotesize Annihilations of $A^{(1)} A^{(1)}$ into Standard Model $Z$ gauge bosons.}
\label{fig:A1A1_ZZ_annihilations_diagrams}
\end{center}
\end{figure}

The annihilation $A^{(1)}$ into $Z$ gauge bosons is mediated only by the SM  Higgs exchange presented in Figure~\ref{fig:A1A1_ZZ_annihilations_diagrams}. The interaction of the $A^{(1)}$ with the Standard Model Higgs boson $h$ is given by
 \begin{equation}
 \mathcal{L}_h^{4D}=-\frac{g_2^2}{2}\frac{(c_w s_{w1}-c_{w1}s_w)^2}{c_w^2}A^{(1)}A^{(1)}h(h+v)
 \end{equation}
where $g_2$ is the $SU(2)$ gauge coupling constant, $s_w$ and $c_w$ are the sine and cosine of the Standard Model Weinberg 
mixing angle ($s_w^2 = 0.23 $) and $s_{w1},~c_{w1}$ are the electroweak mixing angles in the first KK tier and $v \approx 246$ GeV 
is the electroweak scale. The annihilation cross section into a pair of $Z$ bosons reads
 \begin{equation}
 \sigma(A^{(1)}A^{(1)}\rightarrow ZZ) =Y_{A^1}^2 Y_Z^2\frac{s^2-4sm_Z^2 +12m_z^4}{128\pi sm_Z^4(s-m_h^2)^2}\sqrt{\frac{s-4m_Z^2}{s-4m_{A^1}^2}}
 \end{equation}
where $Y_{A^1}= \frac{g_2^2 v (c_w s_{w1}-c_{w1}s_w)^2}{2c_w^2}$ is the Higgs-$A^{(1)}$ coupling and 
$Y_Z = \frac{g_2^2 v}{2c_w^2} $ is the SM Higgs-$Z$ coupling constant.
Expanding the cross section in powers of the relative speed $v_{rel}$ between the $A^{(1)}$ photons  gives
\begin{equation}
v_{rel}\sigma(A^{(1)}A^{(1)}\rightarrow ZZ)\approx a_{ZZ}+b_{ZZ}v_{rel}^2+\mathcal{O}(v_{rel}^4)
\end{equation}
and the first two terms in this non-relativistic expansion are
\begin{eqnarray}
a_{ZZ} =  Y_{A^1}^2 Y_Z^2 \frac{4 m_{A^1}^4-4 m_{A^1}^2 m_Z^2+3 m_Z^4 \sqrt{m_{A^1}^2 - m_Z^2}}{64 \pi m_{A^1}^3 m_Z^4 
(m_h^2 - 4 m_{A^1}^2)^2}
\end{eqnarray}
and
\begin{multline}
b_{ZZ} =- Y_{A^1}^2 Y_Z^2 \times\\
\frac{64 m_{A^1}^8-176 m_{A^1}^6 m_Z^2+4 m_{A^1}^4 (3 m_h^2 m_Z^2+52 m_Z^4)-12 m_{A^1}^2 (2 
m_h^2 m_Z^4+9 m_Z^6)+15 m_h^2 m_Z^6}{512 \pi m_Z^4 (4 m_{A^1}^3-m_{A^1} m_h^2)^3 \sqrt{m_{A^1}^2-m_Z^2}}\,.
 \end{multline}
 
\subsubsection{$A^{(1)}A^{(1)}\rightarrow W^+W^-$}

\begin{figure}[!ht]
\begin{center}
     \includegraphics[scale=0.3]{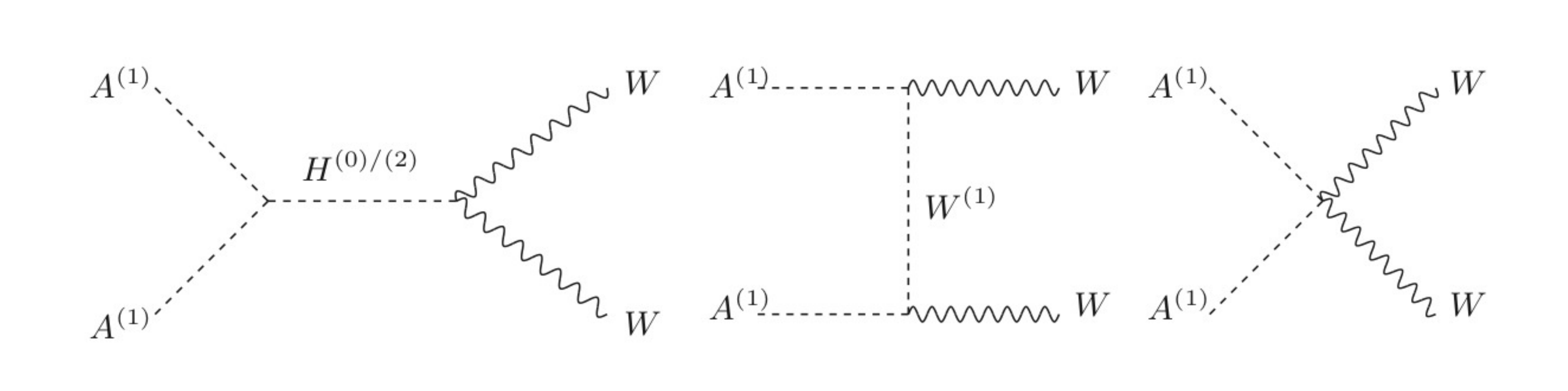}
     \caption{\footnotesize Annihilations of $A^{(1)} A^{(1)}$ into SM W gauge bosons.}
\label{fig:A1A1_WW_annihilations_diagrams}
\end{center}
\end{figure}
The annihilation of $A^{(1)}$ into $W^{\pm}$ gauge bosons is mediated by the SM Higgs exchange in the $s$-channel, by the 
$W^{(1)}$ scalar partners of $W$ in $t$ and $u$-channels and finally by a direct quartic coupling with two Standard Model 
$W^{\pm}$ gauge bosons (see Figure~\ref{fig:A1A1_WW_annihilations_diagrams}). For the annihilation cross section we obtain: 
(here for simplicity we give the results with all the SM masses neglected and for the degenerated KK masses, but the full results are 
retained in the numerical analysis)
 \begin{multline}
  \sigma(A^{(1)}A^{(1)}\rightarrow W^+W^-)  = \frac{g_2 s_{w1}}{2\pi s(s-4m_{A^1}^2)} \times \\
  \left[ \frac{s^3-16sm_{A^1}^4}{\sqrt{s(s-4m_{A^1}^2)}} - 4m_{A^1}^2(s-2m_{A^1}^2)\ln\frac{s+\sqrt{s(s-4m_{A^1}^2)}}{s-\sqrt{s(s-4m_{A^1}^2)}}\right]\,,
 \end{multline}
and the first two coefficients of the non-relativistic expansion are
 \begin{eqnarray}
 a_{WW} &=& \frac{g_2^4 s_{w1}^4}{4\pi m_{A^1}^2}\,,\\
 b_{WW} &=& -\frac{5}{6}a_{WW}\,.
 \end{eqnarray}

\subsubsection{$A^{(1)}A^{(1)}\rightarrow h h$}

\begin{figure}[!ht]
\begin{center}
\includegraphics[scale=0.3]{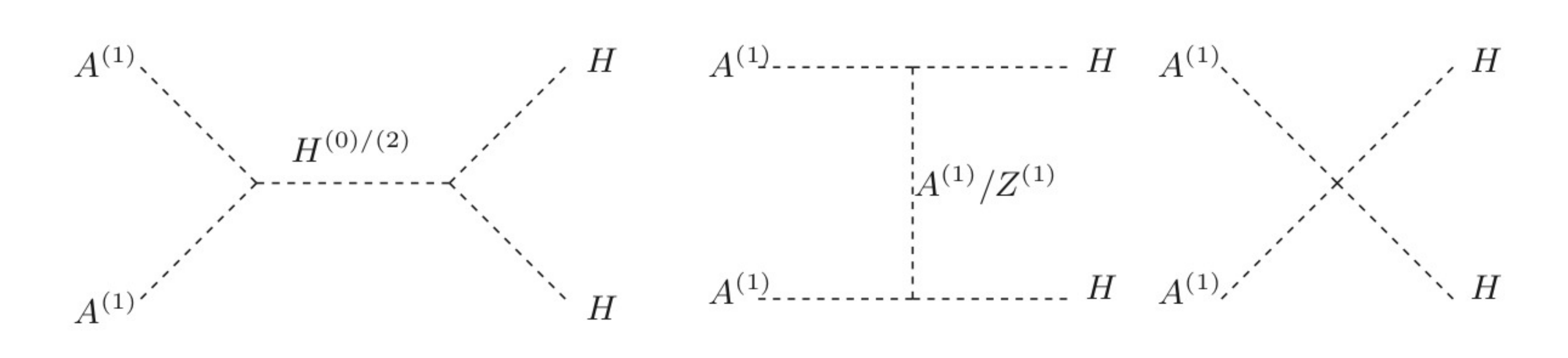}
\caption{\footnotesize Annihilations of $A^{(1)}$ into SM Higgs bosons.}
\label{fig:A1A1_HH_annihilations_diagrams}
\end{center}
\end{figure}

Finally for the Higgs boson production (Figure~\ref{fig:A1A1_HH_annihilations_diagrams}), neglecting all the SM masses, we get

\begin{equation}
 \sigma(A^{(1)}A^{(1)}\rightarrow hh)=\frac{g_w^4 (c_{w1} s_w-c_w s_{w1})^4}{128 \pi c_w^4 \sqrt{s (s-4 m_{A^1}^2)}}\,,
\end{equation}
and the non-relativistic coefficients read:
\begin{eqnarray}
a_{hh} &=& \frac{g_w^4(c_{w1}s_w - c_w s_{w1})^4}{256\pi c_w^4 m_{A^1}^2}\,,\\
b_{hh} &=& = -\frac{1}{2}a_{hh}\,.
\end{eqnarray}

 \subsubsection{$A^{(1)} A^{(1)}\rightarrow f\bar{f}$}

\begin{figure}[!ht]
\begin{center}
\includegraphics[scale=0.3]{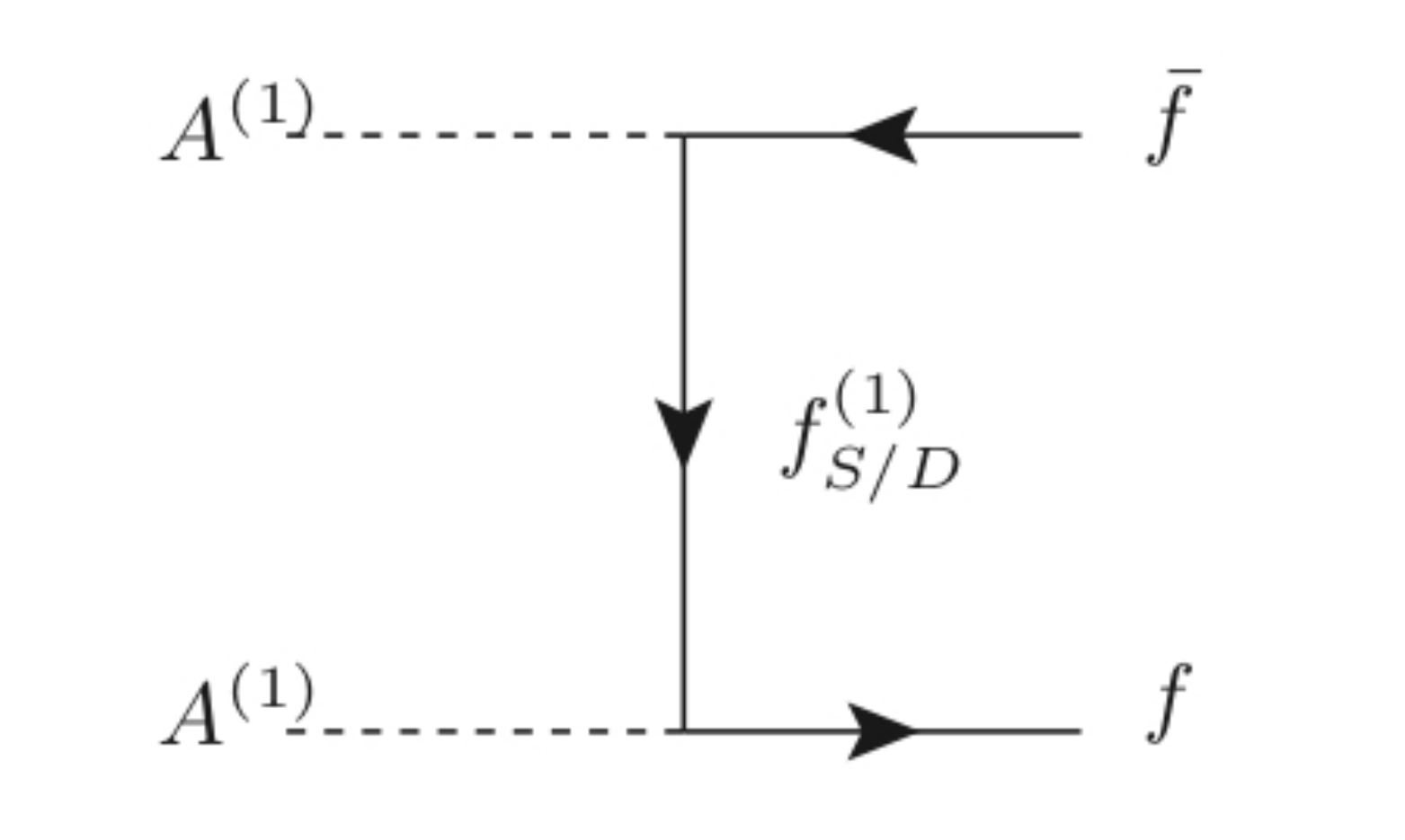}
\caption{\footnotesize Annihilations of $A^1 A^1$ into light SM fermions.}
\label{fig:A1A1_tT_annihilations_diagrams}
\end{center}
\end{figure}

The annihilation of $A^{(1)}$ into light fermionic degrees of freedom $f$ (quarks $u$, $d$, $s$, $c$, $b$ and all the leptons) is 
mediated through the exchange of level one singlet and doublet fermions $f^{1}_{D/S}$ into $t$ and $u$ channels, see Figure~\ref{fig:A1A1_tT_annihilations_diagrams}. Moreover we 
include the Yukawa couplings of the top quark $t\bar{t}h$, thus the annihilation into two top quarks will have additional contribution form 
the SM Higgs in the $s$-channel. The production of neutrinos is mediated only by the exchange of doublet neutrinos $\nu^{(1)}_D$ 
in $t$ and $u$ channel.
The coefficients $a_{FF}$ and $b_{FF}$ are both proportional to the SM fermion mass. In the first approximation they will give both a 
zero result, therefore we give their expressions without neglecting the SM masses but assuming degenerated KK spectrum for 
simplicity:
 \begin{eqnarray}\label{eq:AAff_annihil_analytical}
 a_{FF} &=& (Y_D^2+Y_S^2)^2\frac{m_q^2 (m_{A^1}^2-m_q^2)^{3/2} }{8 \pi m_{A^1}^3 (m_q^2-2 m_{A^1}^2)^2}\,,\\[0.3cm]
 b_{FF} &=& -(Y_D^2 + Y_S^2)^2\frac{m_q^2 \sqrt{m_{A^1}^2-m_q^2}(72 m_{A^1}^6 - 148 m_{A^1}^4 m_q^2 + 82 m_{A^1}^2 m_q^4 - 15 m_q^6) }{192 \pi m^3 (m_q^2-2 m_{A^1}^2)^4}\,,~
 \end{eqnarray}
where $m_{q}$ is the outgoing quark mass and $Y_{S/D}$ are the couplings $f f_{S/D}^{(1)} A^{1}$ defined in Table 
\ref{tab:A1F1f_couplings} in terms of $A=\frac{g_2 s_w c_{w1}}{c_w}$,  $B=\frac{g_2 c_w s_{w1}}{c_w}$ and the fermion electric 
charge $q_f$. For the top quarks one has to include the Yukawa mixings between singlet and doublet states by multiplying the 
corresponding constants by the top mixing angle $\cos\alpha$ for $t^{(1)}_S$ and $\sin\alpha$ for $t^{(1)}_D$.
\begin{table}[!ht]
	\begin{center}
		\begin{tabular}{|c|c|}
\hline
			$~~f_S~~$ & $q_f A$ \\[0.1cm]
			$l_D$ & $\frac{1}{2}(A+B)$ \\[0.1cm]
			$\nu_D$ & $\frac{1}{2}(A-B)$ \\[0.1cm]
			$u_D$ & $\frac{1}{6}A + \frac{1}{2}B$ \\[0.1cm]
			$d_D$ & $\frac{1}{6}A - \frac{1}{2}B$ \\[0.1cm]
\hline
		\end{tabular}
\caption{\footnotesize Couplings of $A^{(1)}$ with fermions.}
\label{tab:A1F1f_couplings}
	\end{center}
\end{table}
The neutrino production cross section expansion will then simply vanish. The top quarks production coefficients will have additional 
contributions from the $s$-channel Higgs exchange.
\begin{eqnarray}
a_{TT} &=& a_{FF}+\frac{g_2^4 (c_w s_{w1}-c_{w1}s_w)^4}{4c_w^4}\frac{m_q^2 \sqrt{m_{A^1}^2-m_q^2}}{8 \pi m_{A^1} 
(m_h^2-4 m_{A^1}^2)^2}\,,\\
b_{TT} &=& b_{FF}- \frac{g_2^4 (c_w s_{w1}-c_{w1}s_w)^4}{4c_w^4}
\frac{m_q^2(24m_{A^1}^4-2m_{A^1}^2(m_h^2+14m_q^2)+3m_h^2 m_f^2)}{64 \pi m_{A^1}(m_h^2-4m_{A^1}^2)^2
\sqrt{m_{A^1}^2-m_q^2}}\,.
\end{eqnarray}

\subsection{Influence of KK mass degeneracy and Higgs contribution}
\label{sec:relab}

\begin{figure}[!ht]
\begin{center}
\includegraphics[scale=0.7]{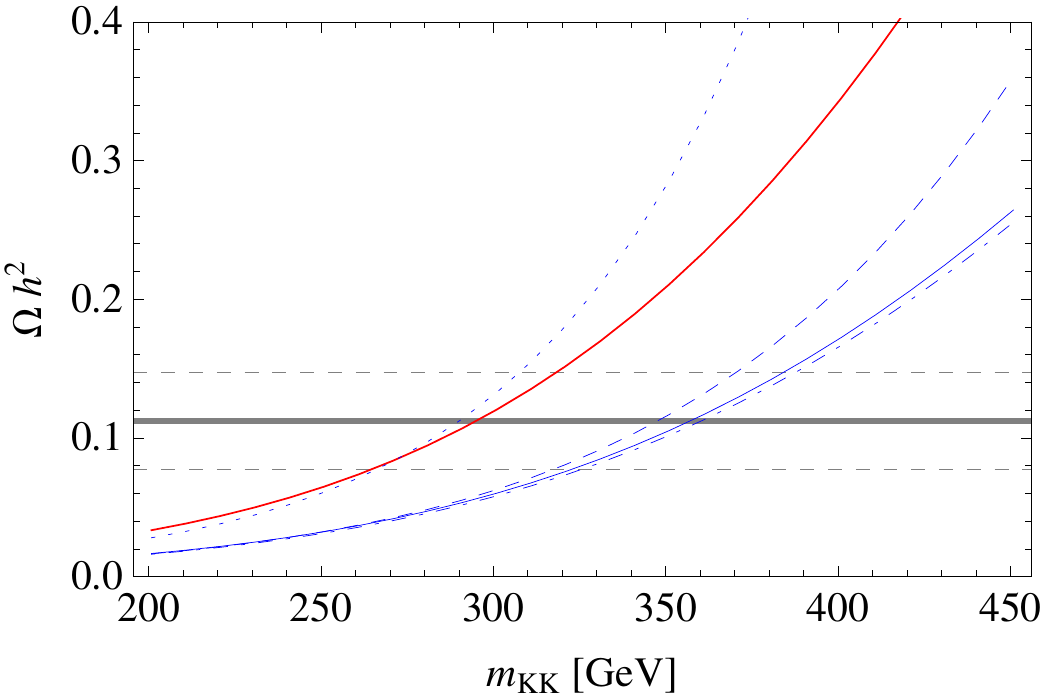}\\
\caption{\footnotesize Analytic result of the relic abundance calculation from annihilations only on the real projective plain. In these plots we show the impact 
of several factors on the relic abundance: we start with the simplest case (blue dotted line) where we assume equal masses
for the first KK level and no SM Higgs exchange in $s$ channel. Then progressively we add the contributions: the $(1)$ level mass corrections (blue dotted), Higgs $s$-channel contribution (blue dot-dashed), relativistic corrections (blue solid line). We add also the same result assuming the symmetric radii $R_4 = R_5$ with all the above corrections included (red solid line).\label{fig:RelicAbundance_plot_annihilations_analytical_v1}}
\end{center}
\end{figure}

In this section we will carefully study the influence of various effects on the relic abundance of Dark Matter:
\begin{enumerate}
	\item degeneracy of masses in the KK tiers,
        \item SM Higgs exchange,
        \item relativistic corrections,
\end{enumerate}
using analytic expressions for cross sections and relic abundance.
The results are 
shown in Figure \ref{fig:RelicAbundance_plot_annihilations_analytical_v1}. We also summarise the allowed values of $m_{KK}$ in Table 
\ref{tab:annihil_mkkbounds_analytic} where we give the preferred ranges for the decoupling limit $R_4 \gg R_5$ on the left and for the 
symmetric degenerate radii $R_4 = R_5$ in the right column. 

\begin{figure}[!ht]
\begin{center}
\includegraphics[scale=0.7]{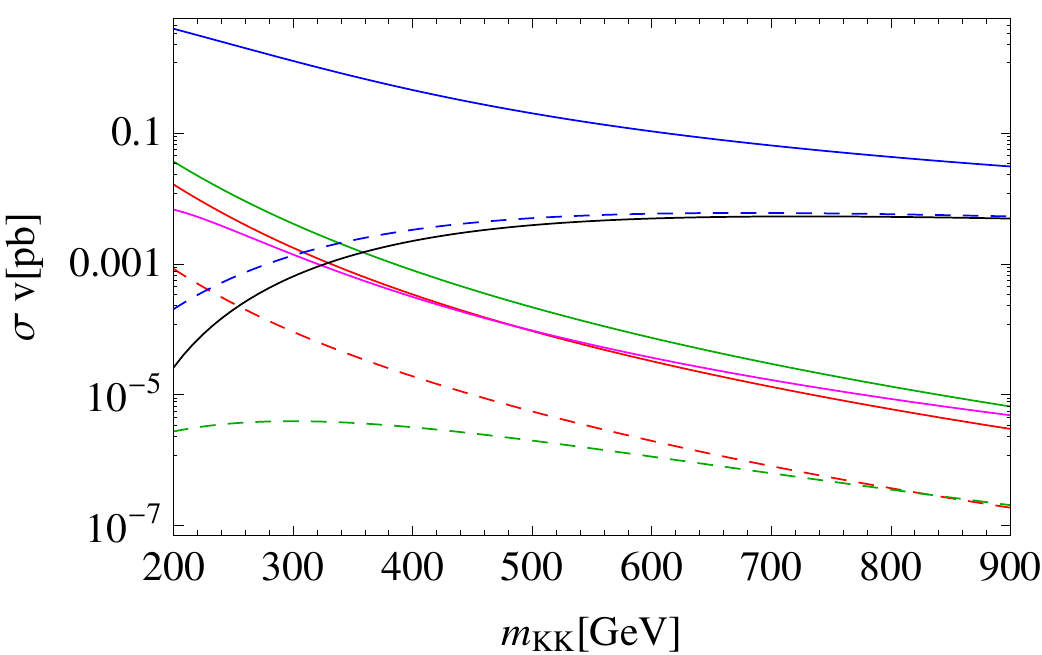}
\caption{Averaged annihilation cross sections $\sigma(A^{(1)} A^{(1)} \rightarrow SM)$ of the LKP into Standard Model final 
states. In red (red-dashed) up (down) type quarks, magenta - tops, green (green-dashed) - charged leptons (neutrinos), blue (blue dashed) - WW (ZZ) gauge bosons, in black - HH. The cross section is calculated at $p_{cms} = 100$ GeV (consistent for 
non--relativistic DM, the maximum of the Boltzmann distribution is given by 
$p_0 = \sqrt{2mkT} = m\sqrt{2k/x_F} \approx 0.28 m_{KK}$ for $k=1,\,x_F=25$ and vary between $56-280$ GeV for 
$m_{KK}=200-1000$ GeV).}
\label{fig:sigmav_annihilations_A1A1}
\end{center}
\end{figure}

\begin{table}
\begin{center}
\begin{tabular}{|c|c|c|}
\hline
 & $R_4 \gg R_5$ & $R_4 = R_5$ \\
\hline
equal masses        & 266 - 307     & 221 - 262\\
corrected masses       & 317 - 370     & 263 - 313\\
+ s-channel Higgs       & 325 - 388     & 267 - 321\\
+ relativistic corrections  & 322 - 384     & 264 - 318\\
\hline
\end{tabular}
\caption{\footnotesize Preferred ranges for $m_{KK}$ (in GeV) from the analytic relic abundance calculation. In the first approximation we include only 
annihilations of LKP and study the impact of including the corrections 1) non-degenerated spectrum of the first KK level (line 2), 2) 
including $s$-channel SM Higgs exchange (line 3), including relativistic correction to the $b-{rel}$ coefficient (line 4). In the first line we 
show the bounds obtained for the simplest case where all the SM masses are neglected and the KK spectrum is fully degenerate. In the 
left column we present the results for the non-degenerated radii $R_4 \gg R_5$, in the right column - for the symmetric radii $R_4 = R_5$. 
\label{tab:annihil_mkkbounds_analytic}}
\end{center}
\end{table}

We first consider only annihilation processes: the average annihilation cross sections as a function of the KK mass are shown in Figure \ref{fig:sigmav_annihilations_A1A1}.

The blue dotted line in Figure 
\ref{fig:RelicAbundance_plot_annihilations_analytical_v1} presents the relic abundance assuming all the $(1)$ level KK states have the same mass and the Higgs couplings are neglected as well. In this regime only annihilations into the $W$ gauge bosons 
contribute considerably. Annihilations into $Z$ and Higgs bosons are turned off and the fermion production is negligible (see 
Figure \ref{fig:sigmav_annihilations_A1A1}). Then we release the assumption of the degeneracy of the KK states (blue dashed line) and use the full one loop spectrum: the relic 
abundance is considerably reduced. If we take a representative point $m_{KK}=350$ GeV, which lies in the range allowed by 
WMAP7 data (for the most complete scenario) we obtain $\Omega h^2 = 0.285$ for the equal mass scenario, a value far above the 
experimental data, which is pushed down to $\Omega h^2 = 0.121$ for the exact spectrum - a value which satisfies the 
experimental bounds. This enormous reduction of about 60\% is due to the considerable changes in the annihilation cross section 
into $W$ bosons. 

\begin{figure}[!ht]
 \begin{center}
 \includegraphics[scale=0.75]{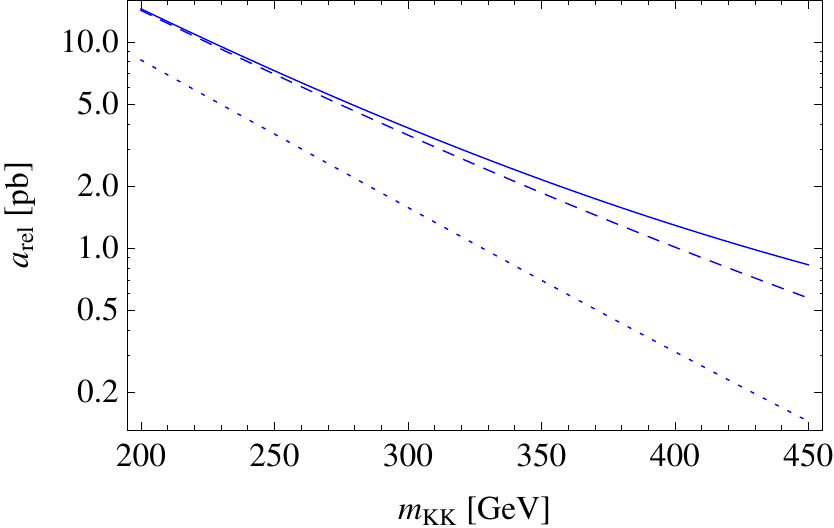}~~~~
 \includegraphics[scale=0.75]{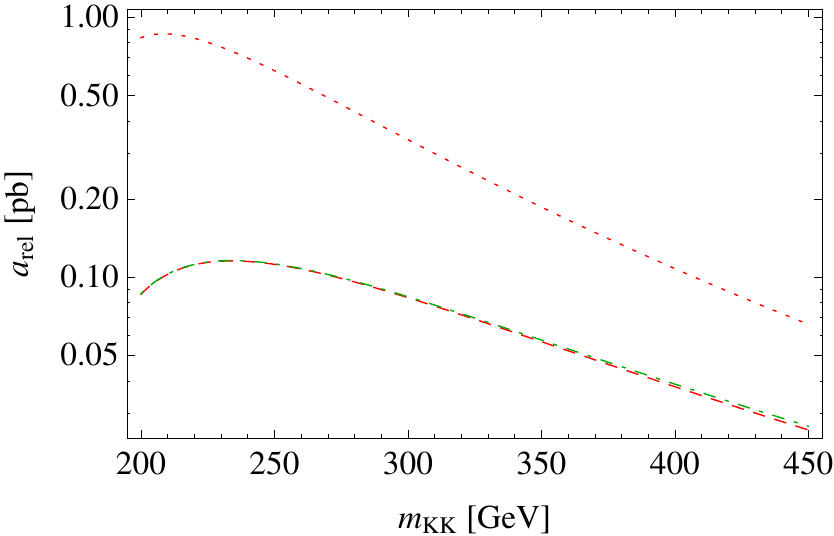}\\
 \caption{\footnotesize First coefficients of the non relativistic expansion $\langle \sigma v_{rel}\rangle \approx a_{rel}+v_{rel}^2b_{rel}$. 
 On the left panel the coefficients of the annihilations into gauge boson are shown, in blue dotted - for equal $KK$ masses, in blue 
 dashed - exact $KK$ masses, in blue dot dashed line - the Higgs contribution. On the right panel - coefficients of the 
 annihilations into all fermions summed. In red dotted - equal $m_{(1)}$ masses, in red dashed - exact $m_{(1)}$ masses 
 and in green we add the Higgs contribution.}
 \label{fig:Arel_annihil_bosons_fermions}
 \end{center}
 \end{figure}

 \begin{figure}[!ht]
 \begin{center}
 \includegraphics[scale=0.8]{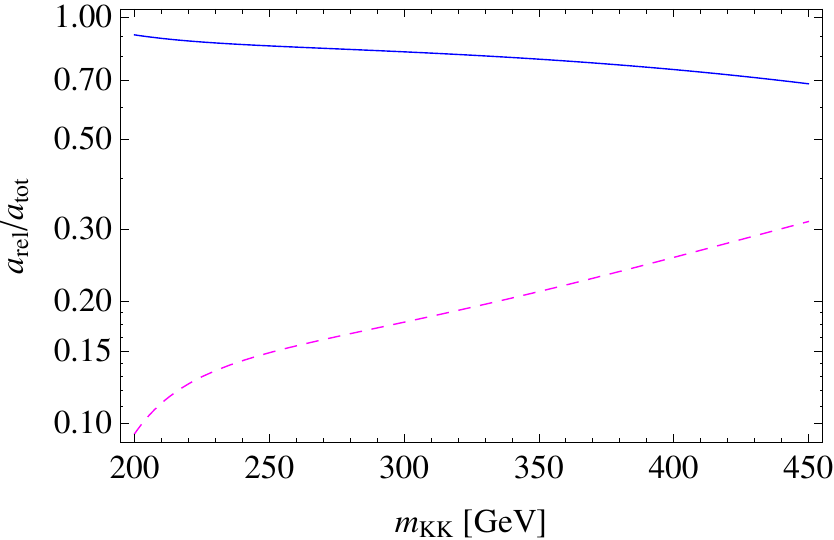}~~~~
 \includegraphics[scale=0.8]{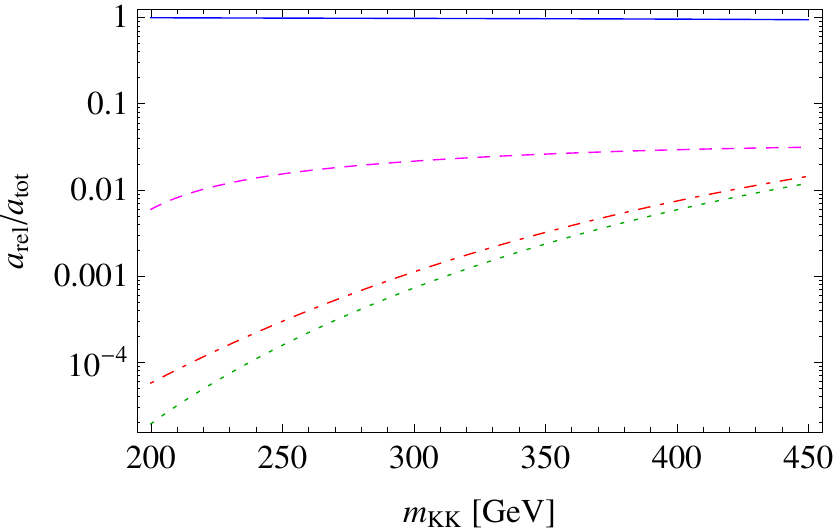}\\
 \caption{\footnotesize On the left: Relative contribution of WW annihilation (blue line) and all fermion annihilations (dashed magenta) for 
 the equal mass spectrum. On the right:  Relative contributions $a_{XX}/a_{tot}$ for the exact spectrum and Higgs channel added. 
 In blue $WW$, in dotted green $hh$, in red dot-dashed $ZZ$, in magenta dashed - $f\bar{f}$ final states.}
 \label{fig:ArelAtot_fermions_W}
 \end{center}
 \end{figure}

As dealing with large analytic formulas is difficult, in Figure \ref{fig:sigmav_annihilations_A1A1} we compare the 
first coefficient of the non--relativistic expansion $a_{rel}$ for the three cases studied here. First note that effectively the contributions 
coming from the annihilations into quark and leptons are almost negligible when compared to the annihilations into gauge bosons which 
are two orders of magnitude larger. Then when we release the equal mass approximation, as the masses of $W^{(1)}$ 
scalars exchanged in the $t$ and $u$ channels appear in the numerator and denominator it is difficult to guess what will be the overall 
impact of the mass degeneracy. But from the plot we see that in the equal mass approximation, the value of the
annihilation cross section $\sigma(A^{(1)} A^{(1)}\rightarrow W^+W^-)$ is significantly reduced: at the sample point $m_{KK}=350$ GeV we find a ratio $a_{WW}^{equal}/a_{WW}^{exact}=0.376$. 
The behaviour of the fermion production is completely different. The result is shown on the right panel of Figure 
\ref{fig:Arel_annihil_bosons_fermions}. We sum the annihilations into all fermionic degrees of freedom and show the first coefficient of 
the non--relativistic expansion for the equal mass spectrum of the first KK level with the red dotted line and for the non-degenerated 
spectrum with the red dashed line. In this case, in contrast to the annihilations into gauge bosons $a_{rel}$, we see that the degeneracy 
of the spectrum will enhance the annihilation cross section $\sigma(A^{(1)} A^{(1)}\rightarrow f\bar{f})$. Analytically this is simple 
to explain as the masses of the first level quarks $q^{(1)}_{D/S}$ exchanged into $t$ and $u$ channels cancel in the numerators while 
taking the trace of the $\mathcal{S}$-matrix element. We are then left with the $m_{q^{(1)}}$ into denominators only. Schematically we 
can write that the annihilation cross section into fermions is proportional to 
\begin{equation}
\sigma \sim \frac{1}{(m_1^2+m_{D/S}^2-m_q^2)^2} \approx \frac{1}{(m_1^2+m_{D/S}^2)^2} = \frac{1}{m_1^4(1+x^2)^2}\; .
\end{equation}
As the quarks receive large loop corrections to the masses, for $m_{KK}=350$ GeV we have $x=\frac{m_D}{m_1}=1.17$ for the 
$t^{(1)}_D$ top quark we obtain
\begin{equation}
\frac{\sigma_{exact}}{\sigma_{equal}} \approx \frac{4}{(1+x^2)^2} = 0.71\,.
\end{equation}
Thus effectively the annihilation cross section into fermions is overestimated if one assumes degenerate KK spectrum.
Contributions coming from quarks being considerably smaller than the gauge boson contribution, the enhancement of the bosonic cross 
section wins and we observe the decrease of the relic abundance. 

Next we add the $s$-channel Higgs exchange, blue dot-dashed line in Figure \ref{fig:RelicAbundance_plot_annihilations_analytical_v1}. The effect is small but visible. As we add new channels to the annihilation cross section 
the relic abundance is reduced. The modification is of about 13\% compared to the previous case where we have considered the 
non-degenerate KK spectrum without Higgs contributions. Note that now all the annihilation channels contribute, that is the cross 
sections $\sigma(A^{(1)} A^{(1)}\rightarrow ZZ,~hh)$ are present. 
The right panel on Figure 
\ref{fig:ArelAtot_fermions_W} shows the relative contributions of different final states into the total annihilation cross section.  The 
$W^+W^-$ gauge boson production still gives the leading contribution, but we can observe the growing contributions of $ZZ$ and $hh$ final 
states. What is worth noticing is that while we do not neglect the Standard Model masses we effectively get new contributions to the total 
annihilation cross section coming from fermionic, $ZZ$ and $hh$ final states. From this analysis we see that these processes cannot 
simply be neglected as their contribution changes of about 13\% the relic abundance prediction.

\section{Relic Abundance - numerical results}
\label{sec:relicAbundanceBounds}

We now turn to studying the RP$^2$ model numerically.
In the following we estimate the WMAP preferred range of the compactification scale $m_{KK}$ of the extra dimensions using the relic abundance 
calculation performed numerically with MicrOMEGAs \cite{Belanger:2001fz, Belanger:2004yn}. The general assumptions are the same as for the analytic calculations, presented at the beginning of section~\ref{sec:analytical_results}.

\begin{figure}
	\begin{center}
 \includegraphics[scale=0.3]{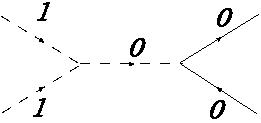}~~~
 \includegraphics[scale=0.3]{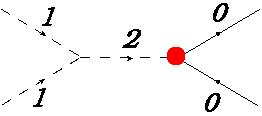}~~~
 \includegraphics[scale=0.3]{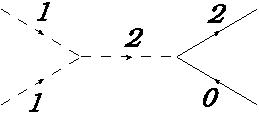}
\caption{\footnotesize Schematic process included in the L2 scenario. The loop couplings marked as red vertexes. }
\label{fig:L2_processes}
\end{center}
\end{figure}

First we have implemented the model into FeynRules \cite{Christensen:2008py} where all the effective Lagrangians of the model in terms of 4-dimensional fields and couplings are given in Mathematica language and interfaced with Monte Carlo generators like CalcHEP \cite{Pukhov:2004ca}. We have implemented the full set of states in tiers $(1)$ and $(2)$ including all bulk couplings and one-loop order masses.
We also implemented the loop induced couplings that mediate the decays of the even states into a pair of SM states: the coefficients are calculated in \cite{Cacciapaglia:2011hx} using the effective counter-terms in the ``magic gauge'' $\xi=-3$, however they are only valid for on-shell external particles.
Therefore, the loop induced couplings can only be used consistently to compute decay widths or processes with a resonant exchange of a (2) state.
A full implementation would require the inclusion of gauge-dependent loop corrections to all vertexes, including bulk ones, and is beyond the scope of this work.
We also modified the implementation in order to reproduce the symmetric radii case ($R_4 = R_5$), in which case the loop corrections to the masses of the even states and their loop induced couplings are corrected by a factor of $2$ and $\sqrt{2}$ respectively.
The presence of two degenerate odd tiers is taken into account by doubling the final result of the relic abundance.
The CalcHEP output from FeynRules was then incorporated into MicrOMEGAs, and we validated the numerical implementation against the analytic results of the previous sections.
Unless otherwise stated, the free parameters of the model are fixed to $m_{loc} = 0$ and $\Lambda R = 10$, while the SM Higgs mass is taken to be $125$ GeV.
The WMAP7 bound on the relic density is $0.0773 < \Omega h^2 < 0.1473$. 
We calculate the relic abundance using the implementation of the model to MicrOMEGAs v2.4.1. 
We study two main cases, in order to understand the impact of various effects on the relic abundance :
\begin{itemize}
   \item L1, where the relic abundance calculation includes only the SM final states $(0)$, i.e. we only consider (co-)annihilation processes 
   $(1)+(1) \rightarrow (0)+(0)$;
   \item L2, where we allow for the even $(2)$ KK modes in the final state, therefore we also consider the processes $(1)+(1) \rightarrow (2)+(0)$.
   
\end{itemize}
In both cases, L1 and L2, the intermediate states are $(0)$ or $(2)$ modes in s-channel or $(1)$ KK modes in t-channel with the couplings at tree level. 
A schematic depiction of the (co-)annihilation processes can be found in Figure \ref{fig:L2_processes}, where we label with a red dot the loop induced couplings.

In each case, we will also show partial results including:
\begin{itemize}
   \item L1A and L2A, where we include only annihilation processes mediated by tree level couplings;
   \item L1AL and L2AL, where only annihilation including loop couplings (s-channel $(2)$ resonances) are considered;
   \item L1C and L2C, with the full set of co-annihilation processes with tree level couplings only;
   \item L1CL and L2CL, where co-annihilation including s-channel $(2)$ resonances are included.
\end{itemize}
For each case, we will study the asymmetric $R_4 \gg R_5$ and symmetric $R_4 = R_5$ cases.
Note that the physically meaningful results are given by the most complete case L2CL.

\subsection{L1 scenario}
First we focus on the L1 scenario where the relic abundance calculation includes final states with only a pair SM states $(0)$.
In general, an odd $(1)$ state can never decay directly into a pair of SM states but will always decay in a lighter $(1)$ state and an SM particle.
Each heavy state produced in (co-)annihilations will therefore undergo a chain decay which will end up with the Dark Matter candidate and SM particles.

\subsubsection{Influence of co-annihilations}

\begin{figure}
	\begin{center}
\includegraphics[scale=0.8]{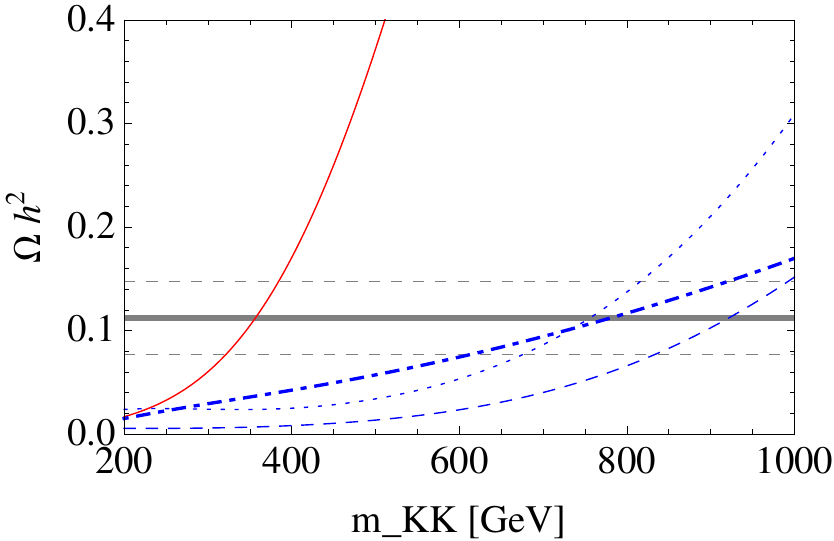}
\includegraphics[scale=0.8]{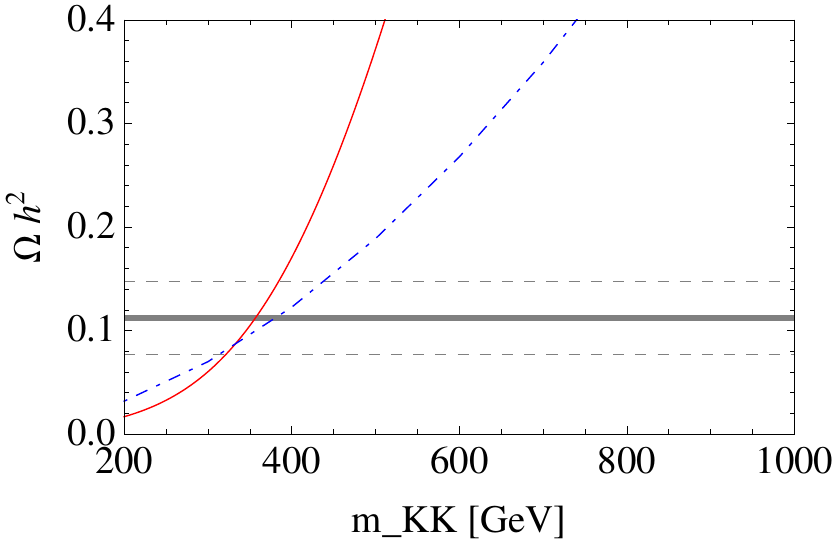}
\caption{\footnotesize On the left. Relic abundance in the asymmetric L1 scenario including co-annihilation with: right-handed leptons (blue dotted), left-handed leptons (blue dashed), and WZ bosons (blue dot-dashed). On the right. Relic abundance including co-annihilations with all the U(1) and SU(2)-only interacting particles (blue dot-dashed); the red line corresponds to annihilations only.}
\label{fig:coannihils_SU2}
\end{center}
\end{figure}

First we examine co-annihilation effects on the relic abundance. 
As it was stated in section \ref{seq:results}, the mass splittings are very small in our model. 
We start by adding the next to LKP particles to co-annihilation channels. In Figure \ref{fig:coannihils_SU2} we show the influence of co-annihilations with: a) right-handed leptons, b) left-handed leptons, and c) electroweak gauge bosons $W^{(1)}$ and $Z^{(1)}$. In all three cases the result is to lower the relic abundance. Right-handed leptons, although having the smallest mass splitting, have weaker interactions than left-handed leptons, which couple also to SU(2) gauge bosons. Therefore the effective cross section with right-handed leptons will be weaker and the thus the relative reduction in the relic abundance smaller. 
 We should notice that for light $m_{KK} \approx 200$ GeV the right-handed leptons induce a small enhancement of the relic abundance, this effect can be explained by very small mass splittings in this region (recall that for $m_{KK}$ GeV the LKP is the right-handed electron) therefore we add only a small contribution to the cross section while the effective number of degrees of freedom increases, becoming $g_{eff}\approx13$ in the $\Lambda = 0$ approximation. 
 For the electroweak gauge bosons, although their interactions are of the same strength as those of left-handed leptons, larger mass splittings will reduce the effect of co-annihilations and the relic abundance will be lager than when including much lighter leptons. 
The result including co-annihilations with all the leptons and $W^{(1)}$ and $Z^{(1)}$ particles is shown in the right panel of Figure \ref{fig:coannihils_SU2}. 

\begin{figure}
	\begin{center}
\includegraphics[scale=0.8]{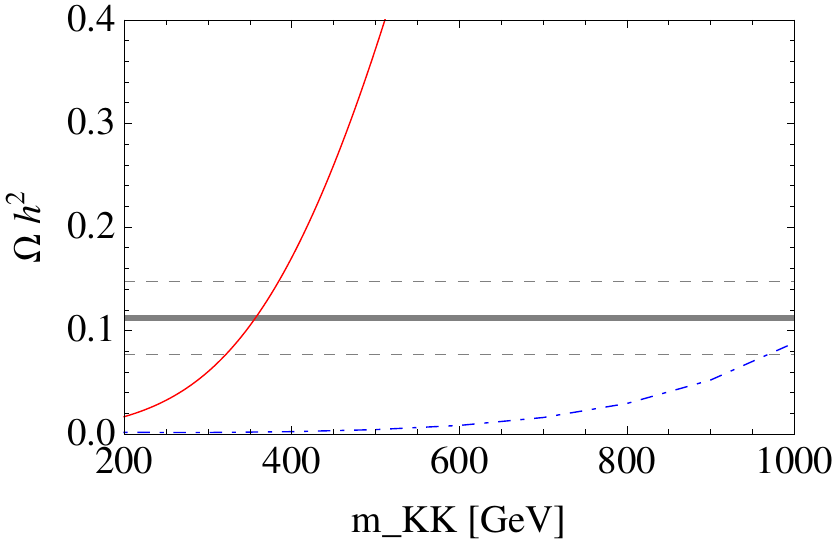}
\includegraphics[scale=0.8]{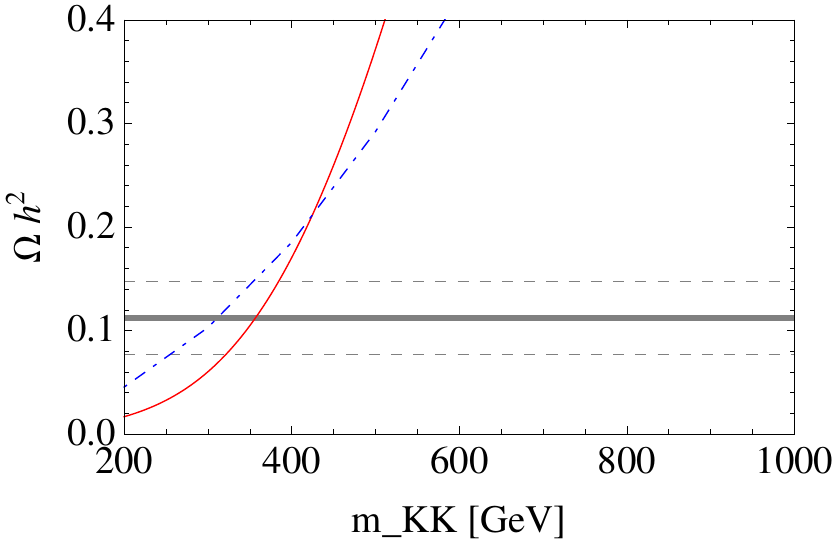}
\caption{\footnotesize On the left: co-annihilations with all the coloured particles in blue dot dashed. On the right: all the possible co-annihilations are included (blue dot dashed line). The red line corresponds to annihilations only.}
\label{fig:coannihils_SU3}
\end{center}
\end{figure}

The influence of adding coloured particles is shown in Figure \ref{fig:coannihils_SU3}. They have the largest mass splitting, however the strong SU(3) interactions enhance the effective cross section and highly reduce the relic abundance. We see that the effect is almost independent of the number of new particles we add: the relic abundance does not change whether we add only singlet quarks, both singlet and doublet quarks, or all quarks together with gauge bosons. This means that the main effect is due to the large value of the annihilation cross section of strongly interacting particles and that, once its addition changes the scale of the effective cross section, the result is not sensitive to the number of states added to the model. In the right panel we show the final result of adding all the possible co-annihilation channels. This plot correspond to the most physical scenario at tree level where we include all the co-annihilations. Note however that it is quite difficult to interpret the enhancement of the relic abundance for low $m_{KK}$ masses as compared to the annihilations only as one would expect the reduction of $\Omega h^2$ for all the range of KK masses. We take therefore the bounds obtained in the case when all the co-annihilations are included as physical bounds $260 <m_{KK} < 355$ GeV in the asymmetric case $R_4 >> R_5$. On the symmetric orbifold the result is simply twice higher relic abundance as we have two independent Dark Matter candidates with degenerate masses and thus the bounds on the KK mass scale are lowered to $m_{KK}< 255$ GeV.

\subsubsection{Influence of the resonances of $(2)$ modes and loop induced couplings}

We study the influence of adding the loop induced couplings. The formal calculation of the effective 
couplings was performed in \cite{Cacciapaglia:2011hx}.
The mass of the $(2)$ KK particles is close to twice the mass of $(1)$ KK particles. Thus, resonant processes in which the 
$(2)$ KK modes propagates in the s-channel are important for the calculation of the relic abundance. Such effects in the mUED model were partially investigated in \cite{Kakizaki:2006dz,Kakizaki:2005uy,Belanger:2010yx}. In the works 
\cite{Kakizaki:2006dz,Kakizaki:2005uy} the second KK particle resonances are studied for the LKP annihilation and 
co-annihilations relevant to the SU(2) singlet leptons, $e^1_S$.  However, it is found that the second KK resonance processes also
play an important role in co-annihilation modes relevant to KK SU(2) doublet leptons and KK Higgs particles. In our model the possible 
resonances can come from the following particles propagating into $s$-channel: $t^{(2)}_{S/D}$, $W^{(2)}$, $Z^{(2)}$, 
$G^{(2)}$, $H^{(2)}$. Other $(2)$ states do not directly couple to a pair of SM particles and thus will not contribute in the co-annihilation processes. In Table \ref{tab:20states_partial_widhts} we give the partial widths of the $(2)$ states to a pair of SM particles.
\begin{table}
\begin{center}
\begin{tabular}{|c|ccc|}
\hline
 & \multicolumn{3}{c|}{Width [GeV] (Br [\%])} \\
$m_{KK}$ [GeV] & 500        & 1000         & 1500 \\
\hline
$Z^5$          & 43\% & 45\% & 43\%\\
$W^5$          & 43\% & 45\% & 43\%\\
$G^5$          & 15\% & 14\% & 14\%\\
$H^5$          & 100\%& 100\%& 100\%\\
$A^5$          & 100\%& 100\%& 100\%   \\
\hline
\end{tabular}
\caption{\footnotesize Branching ratios of the even $(2)$ KK modes into a SM pair.}
\label{tab:20states_partial_widhts}
\end{center}
\end{table}
For the resonant particles mentioned above the corresponding co-annihilation processes are
\begin{eqnarray}
\left.\begin{array}{r}
	t^{(1)}_{S/D}+A^{(1)}/Z^{(1)} \\
	b^{(1)}+W^{(1)}
\end{array}\right\}\rightarrow & t^{(2)}_{S/D} & \rightarrow \textrm{SM}\\
\left.\begin{array}{r}
	A^{(1)}+W^{(1)} \\
	l^{(1)}_D+\nu^{(1)}
\end{array}\right\}\rightarrow & W^{2} & \rightarrow \textrm{SM}\\
	l^{(1)}_{S/D}+\bar{l}^{(1)}_{S/D} \rightarrow & Z^{(2)} & \rightarrow \textrm{SM}\\
\left.\begin{array}{r}
	G^{(1)}+G^{(1)} \\
	q^{(1)}_{S/D}+\bar{q}^{(1)}_{S/D}
\end{array}\right\}\rightarrow & G^{(2)} & \rightarrow \textrm{SM}\\
\left.\begin{array}{r}
	V^{(1)}+V^{(1)} \\
	f^{(1)}_{S/D}+\bar{f}^{(1)}_{S/D}
\end{array}\right\}\rightarrow & H^{(2)} & \rightarrow \textrm{SM}
\end{eqnarray}

Note however that the processes can be suppressed by several factors:
\begin{enumerate}
	\item If the initial particles are heavy, then the process will be Boltzmann suppressed. This condition will be relevant for the initial 
	$G^{(1)}$, $W^{(1)}$, $Z^{(1)}$, $q^{(1)}_{S/D}$ which receive the largest loop corrections to masses.
	\item The BR of the $(2)$ to the SM is small. This condition will reduce the resonant contributions but is much less important than 
	the Boltzmann suppression factors.
	\item As the velocities of the particles near the freeze out temperature are non-relativistic, if one is far below the resonance 
	condition, that is the inequality  $m_{(1)}+m'_{(1)}\ll m_{(2)}$ holds, then the process will be Boltzmann suppressed as well due to large 
	momenta of the incoming particles required to produce the resonance.
\end{enumerate}
We define $m_{KKres}$ the KK mass at which the resonant condition
\begin{equation}\label{eq:res_cond}
	m_{(1)}+m'_{(1)}= m_{(2)}
\end{equation}
is verified. Among many of the kinematically allowed resonances listed above, the contributions of some of them will be highly reduced:
\begin{enumerate}
	\item $t^{(2)}_{S/D}$ have low branching ratios into SM particles and the initial states producing this resonance are heavy, thus these processes will be both Boltzmann and BR suppressed.
	\item $Z^{(2)}$ and $W^{(2)}$ resonances have order $1$ BRs to SM particles. The initial states $l^{(1)}$ however are quite light and the process will be Boltzmann suppressed as we are far below the resonance condition.
        \item $G^{(2)}$ also has sizeable BRs into SM states. However the $(1)$ modes producing the resonance are the particles with the largest mass splittings in the model, therefore the processes will be strongly suppressed by Boltzmann factors.
	\item $H^{(2)}$ particles, in spite of very weak couplings are produced by all the $(1)$ co-annihilating states including $A^{(1)}$, therefore the effective cross section will be enhanced mainly due to the process 
	$A^{(1)} + A^{(1)} \rightarrow H^{(2)} \rightarrow \textrm{SM}$. This process will also be very sensitive to the mass of $H^{(2)}$, which is controlled by the free parameter $m_{loc}$.
\end{enumerate}

\begin{figure}[!ht]
\begin{center}
\includegraphics[scale=0.7]{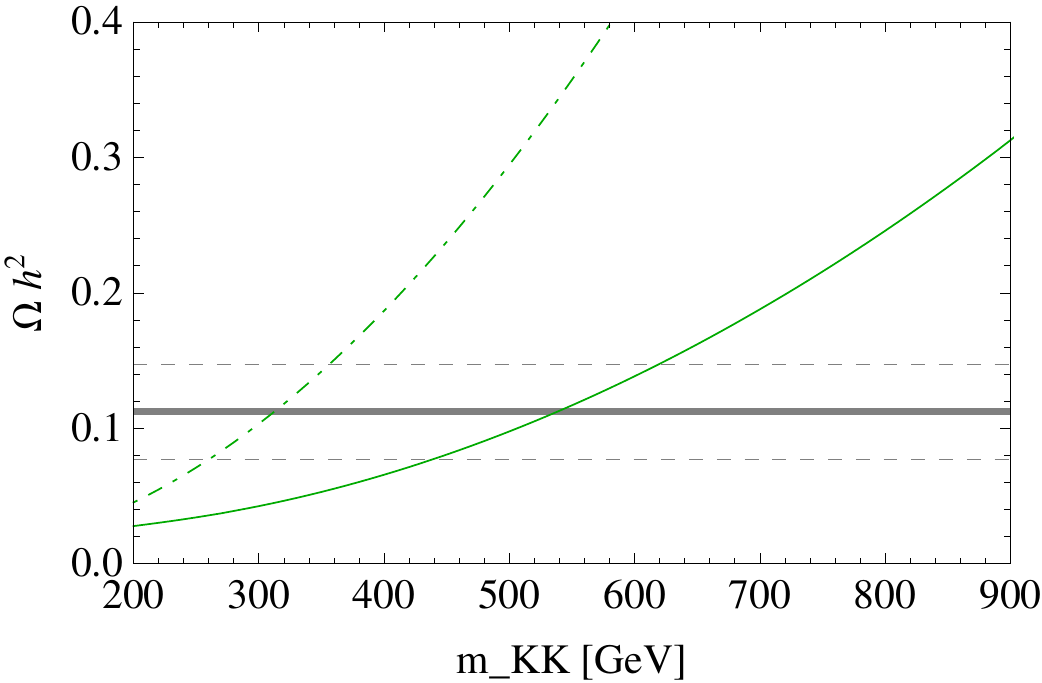} 
\includegraphics[scale=0.7]{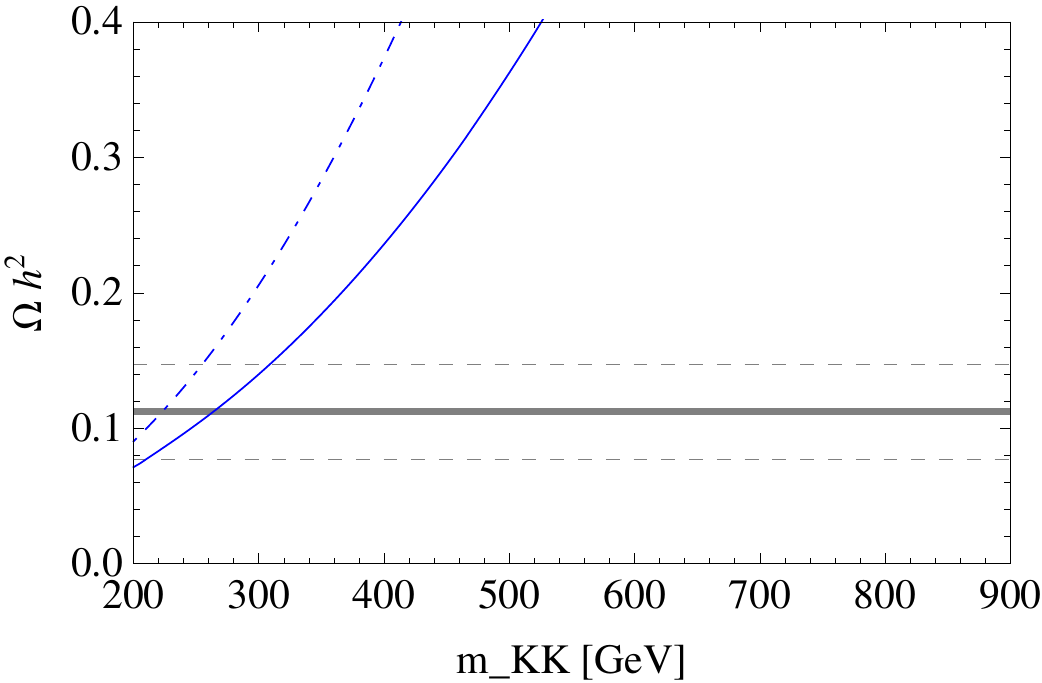}
\caption{\footnotesize Relic abundance as a function of $m_{KK}$ in the L1 scenario. In both plots we use the convention: 
dot-dashed lines for co-annihilations (L1C); solid line for co-annihilations + resonances (L1CL). In green (left panel) the asymmetric $R_4 \gg R_5$ case, in 
blue (right) the symmetric one $R_4 = R_5$.}
\label{fig:L1_relicAbundance}
\end{center}
\end{figure}

The numerical results in the asymmetric (left panel) and symmetric (right panel) cases in L1 scenario are shown in Figure \ref{fig:L1_relicAbundance}: we can immediately see that the loop mediated resonances play a crucial role in the relic abundance calculation.
In the symmetric case, the effect seems to be smaller: this is due to the fact that the divergent loop corrections to the masses of $(2)$ states are larger, therefore many of the resonances are too heavy to be produced in highly non-relativistic collisions of lighter KK modes.

The effect of including loop induced couplings (tier $(2)$ resonances) when only annihilations are taken into account is quite small. The only resonant particle 
present in annihilation processes is the Higgs $H^{(2)}$ in the reaction $A^{(1)} A^{(1)} \rightarrow t\bar{t}$. The resonant condition $m_{A^{(1)}} = m_{H^{(2)}}$ is reached from below: for 
$m_{KK}\leq m_{KKres}=267$ GeV the condition $2m_{A^1}< m_{H^5}$ holds therefore we are below the resonance and heavy Higgs can effectively enter as a resonant state lowering the relic abundance. Once the resonant value is exceeded $m_{KK} > m_{KKres}$ we enter in the regime above the resonance and thus resonant contributions stop. 
The effect of Higgs resonance can only be seen when only annihilations are taken into account. When we include co-annihilations, all the resonant particles mentioned above contribute. The condition \ref{eq:res_cond} 
can be met for the processes 

\

\begin{tabular}{ll}
	$t^{(1)}_{S/D}+A^{(1)}/Z^{(1)}\rightarrow t^{(2)}_{S/D}$ & at\; $m_{KKres}\approx 385/680$ GeV, \\[0.3cm]
         $W^{(1)}+A^{(1)} \rightarrow W^{(2)}$ &at\;  $m_{KKres}=277$ GeV, \\[0.3cm]
         $t^{(1)}_{S/D}+t^{(1)}_{S/D} \rightarrow G^{(2)}$ &at\; $m_{KKres}=740$ GeV.  \\[0.7cm]
\end{tabular}

In all the above processes, the resonant condition is achieved from above, i.e. for $m_{KK}\leq m_{KKres}$ we the inequality $2m_{(1)} > m_{(2)}$ holds and the resonances will be turned 
on for $m_{KK}$ higher than the threshold values. 
The result of this is a strong reduction of the relic abundance for high $m_{KK}$ 
masses (compare dot-dashed and solid lines in Figure \ref{fig:L1_relicAbundance}). The co-annihilations into $H^{(2)}$ always pass through the resonant condition from below and contribute at low masses but the minimum corresponding to $A^{(1)} A^{(1)} \rightarrow H^{(2)}$ spread out by stronger contributions of $W^{(2)}$. The main contribution of about 30\% comes although from the $W^{(2)}$ gauge boson as its decay width into SM particle is 
larger than for other $(2)$ states. The $(2)$ gluons have large decay rates into two SM states but due to their large mass splittings 
their influence will be strongly suppressed by the Boltzmann factor.

Note finally that here we always fix $m_{loc}=0$ GeV, and we will study the dependence of the resonance on this parameter in a following dedicated section.

\subsection{L2 scenario}

In general, an even $(2)$ state can decay into a pair of odd $(1)$ states, into a lighter $(2)$ state plus a SM state or into a pair of SM states.
All such final states give comparable partial widths: the first two are mediated by tree level couplings however they are phase space suppressed by the small loop-induced mass splittings (the decays are nearly on-threshold), while the latter is mediated by a loop induced coupling.
Each state will therefore undergo a chain decay which will end up with only SM particles if a decay via loop couplings is finally met, or into a pair of Dark Matter candidate if a decay into a pair of odd states is met.
In the former case, the annihilation into a $(2)$ state will contribute to the annihilation into SM states.
In Table \ref{tab:20BrtoSM_incl} we list the inclusive branching ratio into SM states, which takes into account the full decay chains.
As we can see, the lighter states (in particular $A^{(2)}$, $H^{(2)}$ and leptons), have a 100\% decay rate into SM final states: this is due to the smallness of the loop corrections that strongly suppresses or closes the phase space of the decays into odd states (the situation for the Higgs may change for large and positive $m_{\rm loc}^2$, which will significantly increase its mass).
Therefore, in the following numerical calculations we will consider all the level-$(2)$ modes in the final state as SM states, thus contributing to the (co-)annihilation cross sections.  
\begin{table}
 \begin{center}
 \begin{tabular}{|c|cc|}
 \hline
    & \multicolumn{2}{c|}{Inclusive BR into SM} \\
 $m_{KK}$ [GeV] & 500 & 1000 \\
 \hline
  $l^{(2)}_S$ & 100\% & 100\%  \\
  $l^{(2)}_D$ & 100\% & 100\%  \\
  $q^{(2)}_S$ & 87\% & 86\%  \\
  $q^{(2)}_D$ & 58\% & 54\%  \\
 $t^{(2)}_S$ & 87\% & 80\%  \\
 $t^{(2)}_D$ & 60\% & 52\%   \\
 $Z^{(2)}$   & 60\% & 60\%  \\
 $W^{(2)}$   & 60\% & 60\%  \\
 $G^{(2)}$   & 38\% & 51\%   \\
 $H^{(2)}$   & 100\%  & 100\%  \\
 $A^{(2)}$   & 100\%  & 100\%  \\
 \hline
 \end{tabular}
 \caption{Inclusive Branching Ratios of $(2)$ KK modes into final states with only SM particles.}
\label{tab:20BrtoSM_incl}
 \end{center}
 \end{table}
 \begin{figure}
\begin{center}
\includegraphics[scale=0.7]{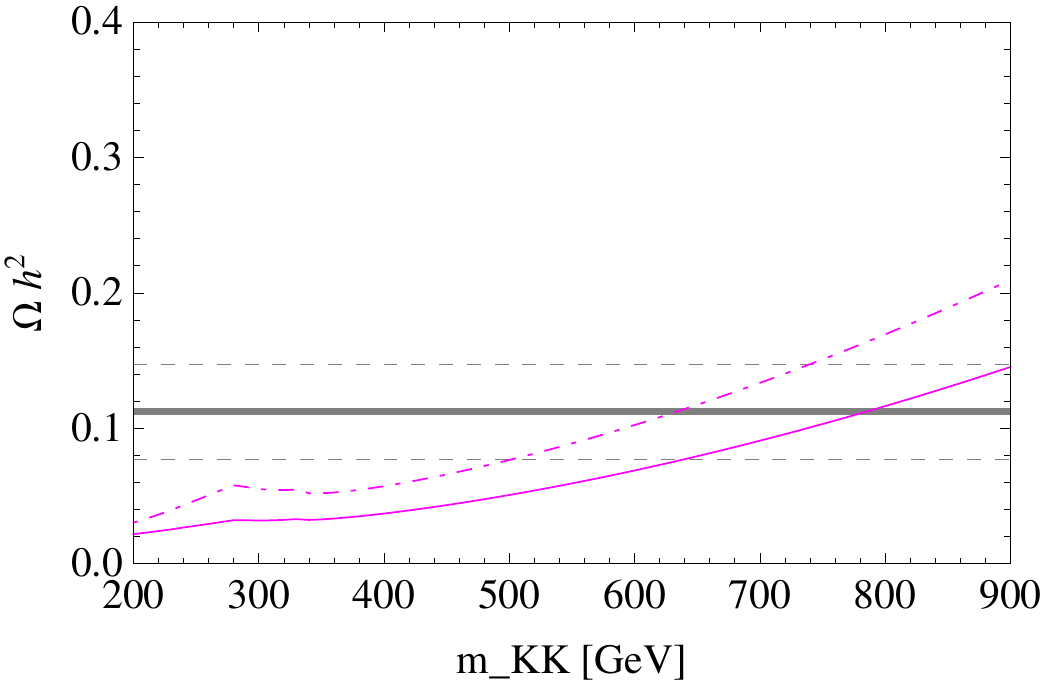}
\includegraphics[scale=0.7]{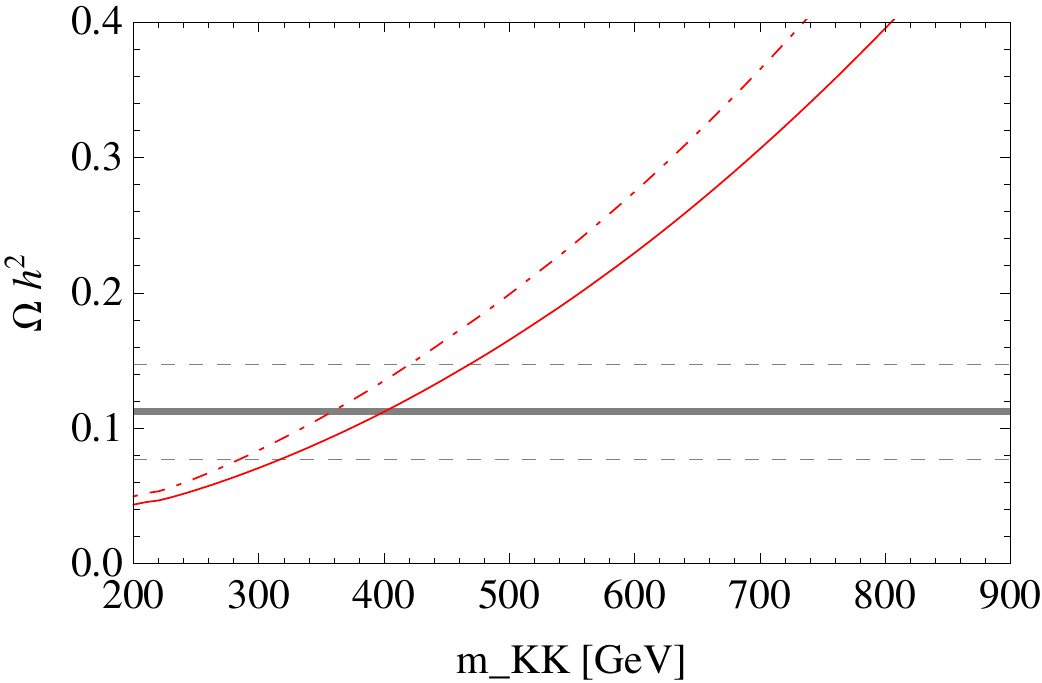}
\caption{\footnotesize Relic abundance as a function of $m_{KK}$ in the L2 scenario. In all the plots we use the convention: dot-dashed lines for co-annihilations only (L2C), 
solid line for co-annihilations + resonances (L2CL). In magenta (left panel) the asymmetric case $R_4 \gg R_5$, in red (right panel) the symmetric one $R_4 = R_5$.}
\label{fig:L2_relicAbundance}
\end{center}
\end{figure}\begin{figure}[!ht]
\begin{center}
\includegraphics[scale=0.7]{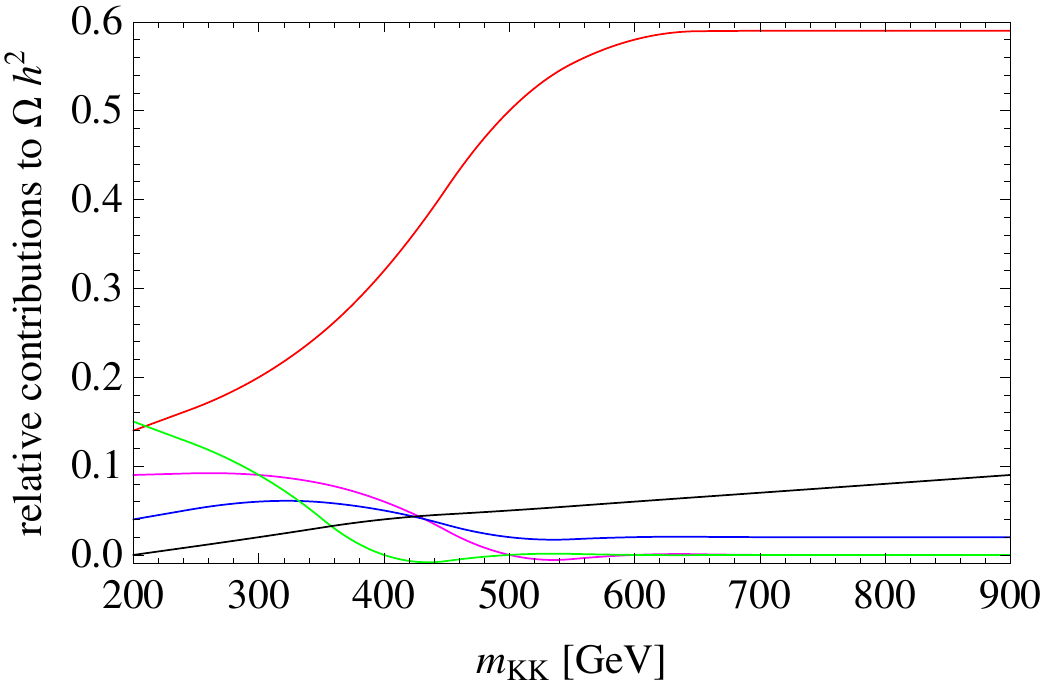} 
\includegraphics[scale=0.7]{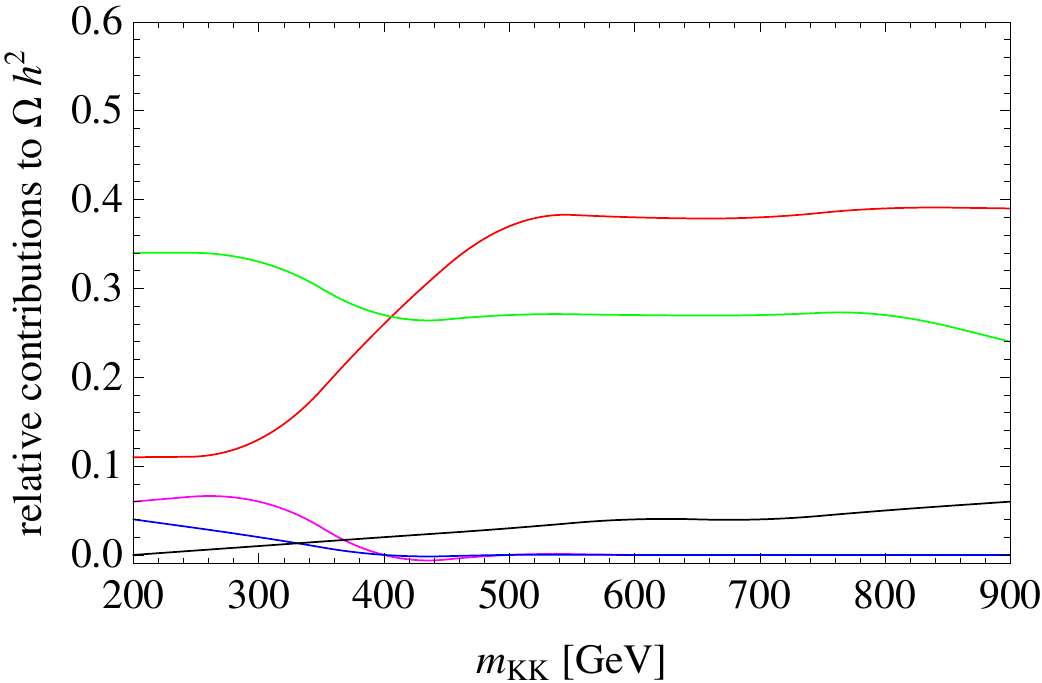}
\caption{\footnotesize Relative contributions of the partial annihilation cross sections to the relic abundance in L2 scenario when all the 
co-annihilation channels are open. On the left panel we see the dominant contribution of $l^{(1)}_{S/D} A^{(1)}\rightarrow l A^{(2)}$ in red . Other 
channel are $l^{(1)}_{D} \nu^{(1)}\rightarrow q\bar{q}$ co-annihilation in green, $l^{(1)}_{D} \nu^{(1)}\rightarrow A H^{(2)}$ in magenta, $q^{(1)} 
A^{(1)}\rightarrow q A^{(2)}$ in blue and $b^{(1)}_{D} W^{(1)}\rightarrow b W^{(2)}$ in black. On the right panel the same colours are applied for the 
contributions when we include the loop couplings ($(2)$ resonances). We see the importance of the $W^{(2)}$ resonance in the 
$l^{(1)}_{D} \nu^{(1)}\rightarrow q\bar{q}$ processes when the loop couplings are turned on.}
\label{fig:L2_relativecontributions}
\end{center}
\end{figure}
 
 In the L2 scenario the bounds do not change, compared to L1 case, if we include only annihilations. This is reasonable as the 
main processes that contribute in both cases are the annihilations $A^{(1)} A^{(1)} \rightarrow W^+ W^-,ZZ,HH$ which are mediated only by 
tree level couplings. For kinematic reasons, the even $(2)$ states cannot be produced from two $A^{(1)} A^{(1)}$.

Including co-annihilations we have many new final states including $A^{(2)}$ and $H^{(2)}$ as compared to the L1 scenario. Analysis of partial co-annihilation processes present similar behaviour to the L1 scenario therefore here we consider only the final result when all the co-annihilations are taken into account.
Due to many new channels including $(2)$ states, as it can be seen form Figure \ref{fig:L2_relicAbundance}, the relic abundance bounds on $m_{KK}$ will be pushed up to 
$m_{KK}\approx800$ GeV in the asymmetric case and to $m_{KK}\approx400$ GeV in the symmetric orbifold. The main contributions are shown in Figure \ref{fig:L2_relativecontributions}.

\subsection{Comparison of $m_{KK}$ bounds in L1 and L2 scenarios}

\begin{figure}[h]
\begin{center}
\includegraphics{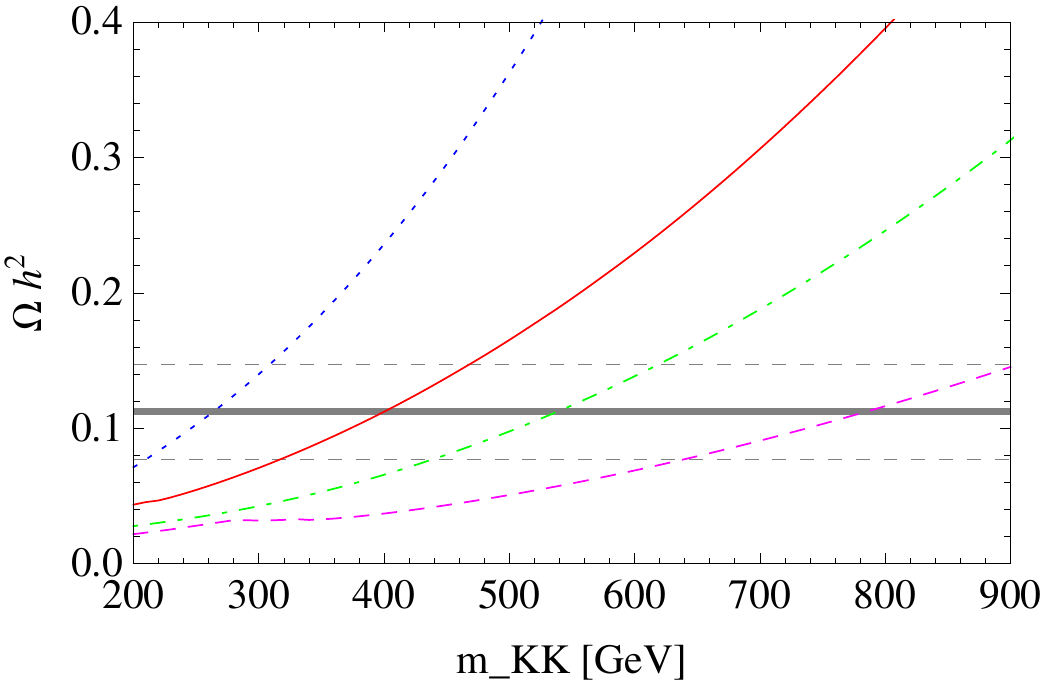}
\caption{\footnotesize Relic abundance as a function of $m_{KK}$, after taking into account all the co-annihilations and loop couplings. Form top to 
bottom we have: blue dashed symmetric L1 $R_4 = R_5$, red solid symmetric L2 $R_4 = R_5$, green dot-dashed asymmetric L1 $R_4 \gg R_5$, magenta dashed asymmetric L2 $R_4 \gg R_5$.}
\label{fig:L1_L2_bound_summary1}
\end{center}
\end{figure}

In Table \ref{tab:L1_L2_bound_summary1} we summarise the bounds on the $m_{KK}$ mass scale deduced from the full (including 
all the co-annihilation channels) relic abundance calculation in the two scenarios L1 and L2 using MicrOMEGAs, while a plot of the relic abundance as a function of $m_{KK}$ is in Figure \ref{fig:L1_L2_bound_summary1}.
The loop induced couplings do not influence the relic abundance bounds when we open only annihilation channels. 
The only resonant particle in this case is the heavy Higgs $H^{(2)}$ which produces a resonant minimum but for the $m_{KK}$ far below 
the expected range compatible with the WMAP data. This behaviour is evident in both scenarios L1 and L2, moreover we have the same 
bounds for $m_{KK}$ with the expected value set quite low at 360 GeV in the asymmetric orbifolds $R_4 \gg R_5$ and at 290 GeV in 
the symmetric case $R_4 = R_5$. This estimation gives the first hint of the possible KK scale of extra dimensions but the fully consistent physical result must contain all co-annihilations and loop couplings.

\begin{table}[!ht]
\begin{center}
\begin{tabular}{|c|c|c|}
\hline
model       & L1            & L2 \\
       & $m_{KK}$[GeV] & $m_{KK}$[GeV] \\
\hline
$R_4 \gg R_5$ & 440 - 620     & 640 - 900        \\
$R_4 = R_5$   & 210 - 310     & 315 - 470       \\

\hline
\end{tabular}
\caption{\footnotesize Preferred ranges for $m_{KK}$ from the relic abundance calculation in the models L1 and L2 for asymmetric radii $R_4 \gg R_5$.}
\label{tab:L1_L2_bound_summary1}
\end{center}
\end{table}

In the L1 scenario when we include all the co-annihilation channels but not loop induced couplings the preferred $m_{KK}$ values lie  in the range $260 < m_{KK} < 355$ GeV in the asymmetric case and $ m_{KK} < 255$ GeV in the symmetric orbifold. When we add the loop couplings the resonant contributions lower the $\Omega h^2$ value of 52\% and the expected mass scale is pushed up to $440 < m_{KK} < 620$ GeV for $R_4 \gg R_5$ ($210 < m_{KK} < 310$ GeV for $R_4 = R_5$).
In the L2 scenario, where the $(2)$ states are allowed in the final states, the impact of co-annihilations is more important which translates into higher mass scales of $500 < m_{KK} < 740$ GeV for $R_4 \gg R_5$ ($285 < m_{KK} < 420$ GeV for $R_4 = R_5$) correspondingly. 
Adding loop couplings, the $\Omega h^2$ value decrease further of about 36\% (15\%) with respect to the co-annihilations only setting the mass range at $640<m_{KK}<900$ GeV ($315<m_{KK}<470$ GeV for $R_4 = R_5$).

The relative impact of opening the $(2)$ final states is always to decrease the relic abundance value and in our study we observe a 
reduction of about 50\% (45\%) between the L1 and L2 scenarios when all the co-annihilation channels and the loop induced couplings are 
taken into account. The physical bounds set by the full one loop calculation are summarised in Table \ref{tab:L1_L2_bound_summary1} and the relic abundance as a function of $m_{KK}$ is shown in Figure \ref{fig:L1_L2_bound_summary1}.

\section{Cut-off dependence of the relic abundance}
\label{sec:cutoff}

In this section we are interested in the dependence on the cut-off of the relic abundance calculation.
The spectrum of the model as well as the loop induced couplings are logarithmically sensitive to the cut-off of the effective extra dimensional theory. The numerical impact can be seen from Table \ref{tab:mass_splittings}, where we show 
the mass splitting for $m_{KK}=500$ and $800$ GeV for two different values of the $\Lambda R$ parameter, and in Figure \ref{fig:Mass_Splittings_variations_with_LamR}.
\begin{figure}[!ht]
\begin{center}
\includegraphics[scale=0.7]{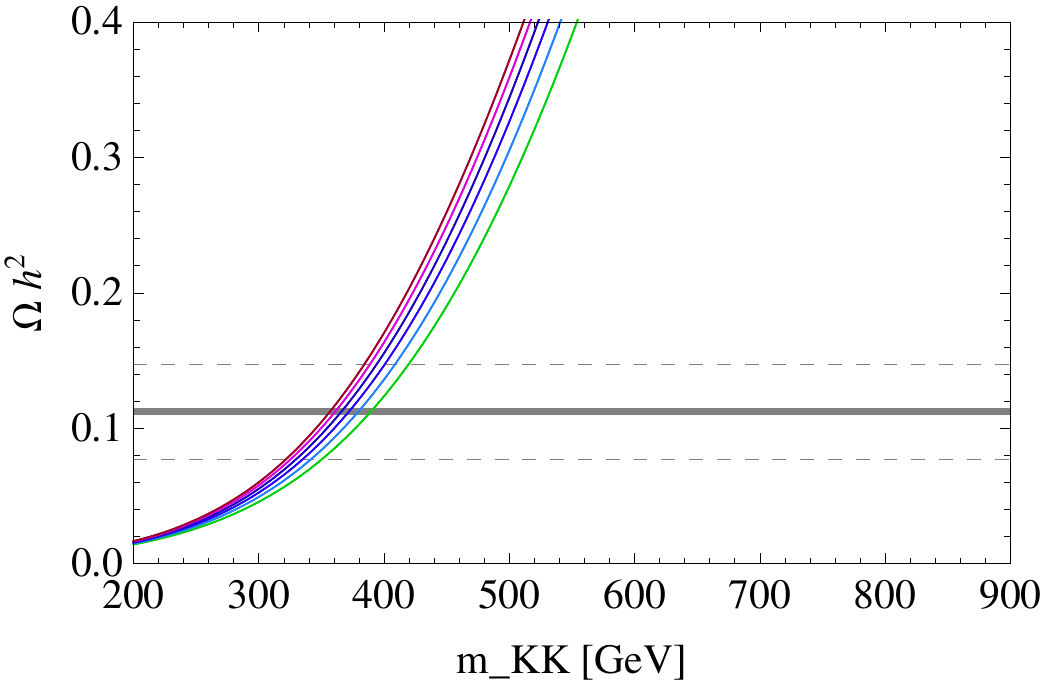}
\includegraphics[scale=0.7]{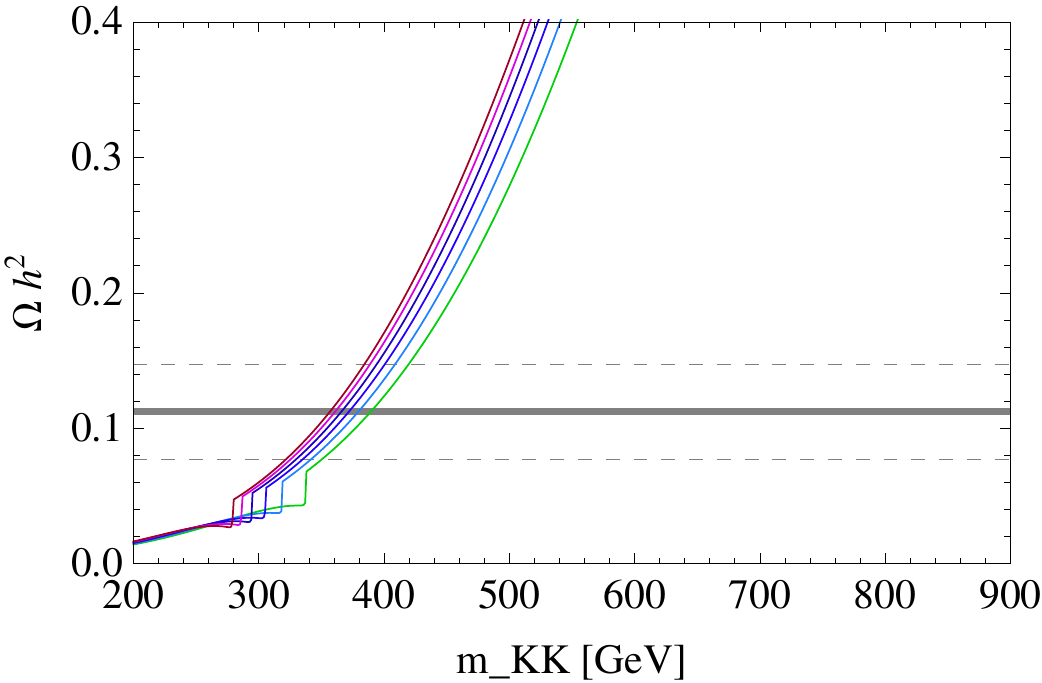}\\
\includegraphics[scale=0.7]{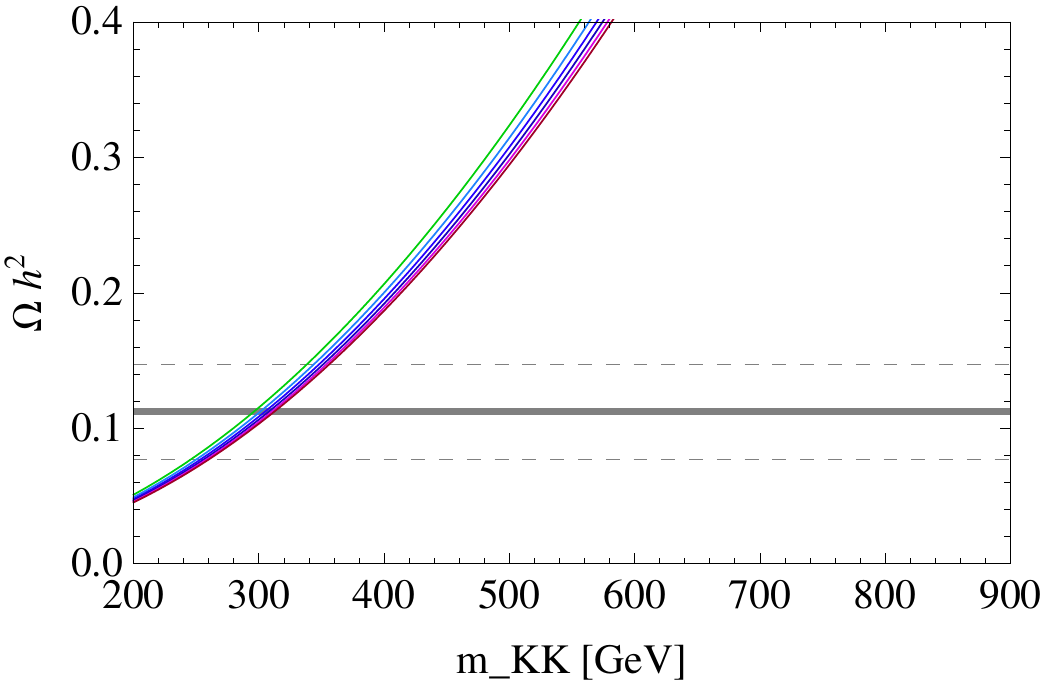}
\includegraphics[scale=0.7]{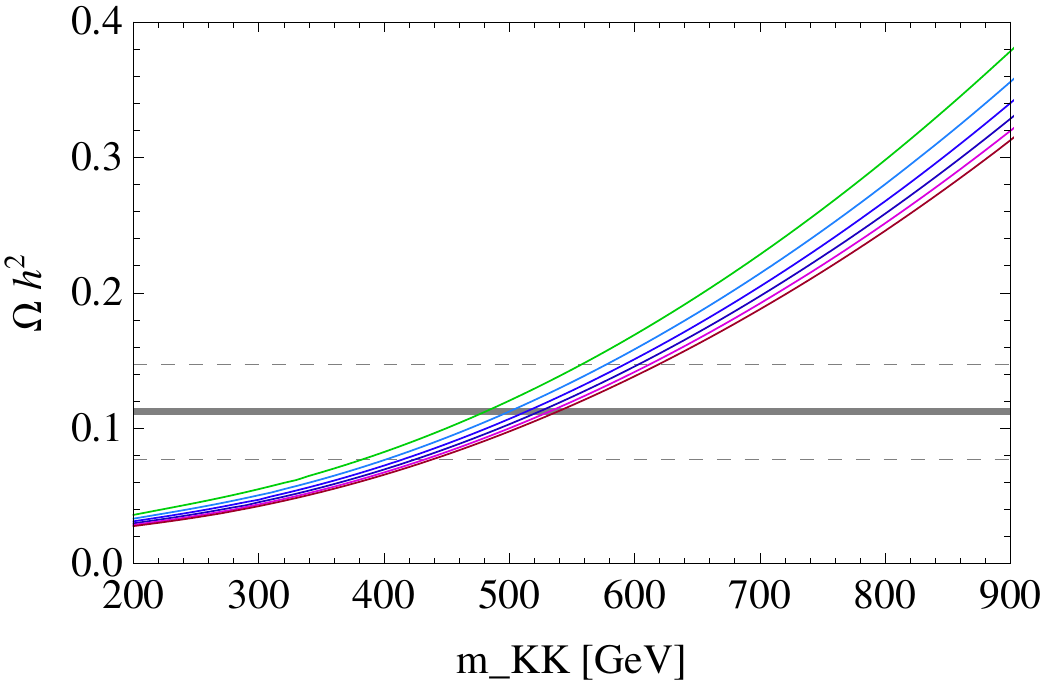}
\caption{Variations of the relic abundance for $R_4 \ll R_5$ in L1A (top left) and L1AL (top right), L1C (bottom left) and L1CL 
(bottom right) with the cut-off. Going from magenta to green the cut off is lowered $10 > \Lambda R > 5$. In black: $\Lambda R = 10$ as 
for every other calculation where we not vary the cut-off.}
\label{fig:L1_OMH2_LRscan}
\end{center}
\end{figure}

The numerical results for the L1 scenario with asymmetric radii can be found in Figure \ref{fig:L1_OMH2_LRscan}, and very similar behaviour can be seen for symmetric radii and in L2.
When we include only annihilations, the growing cut-off has the effect of increasing the relic abundance (top panels in Figure \ref{fig:L1_OMH2_LRscan}). In contrast, when we include all the co-annihilations the situation is reversed - with growing cut off $\Omega h^2$ is reduced (bottom panels in Figure \ref{fig:L1_OMH2_LRscan}). 
To understand these effects we need to know first of all which quantities are most influenced by the cut-off. 
Of course the principal 
influence will be on the mass spectrum of the particles which is explicitly cut-off dependent. Then, while we include loop induced 
processes, the effective couplings violating the KK number are also dependent on the cut off scale of the theory.

\begin{figure}
\begin{center}
\includegraphics[scale=0.7]{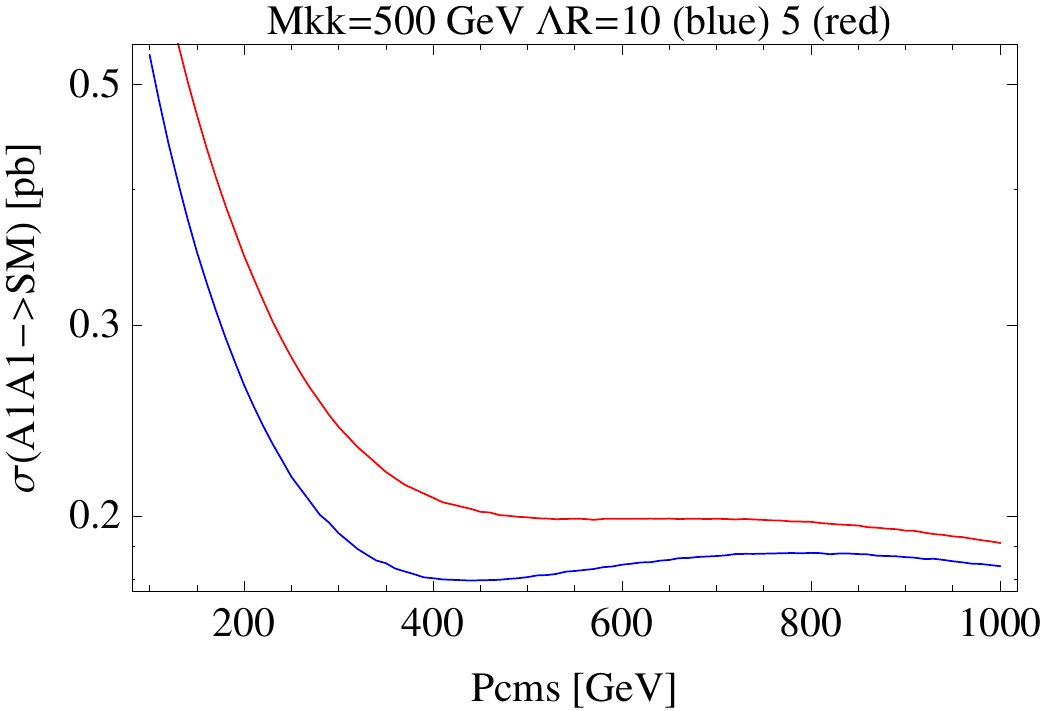}
\caption{Total annihilation cross section $A^{(1)} A^{(1)} \rightarrow SM$ as a function of $p_{cms}$ for $M_{KK} = 500$ GeV 
and $\Lambda R = 5,~10$ in red and blue respectively.}
\label{fig:A1A1_annihilation_xsubsections_with_pcms_Mkk500_LR_5_10}
\end{center}
\end{figure}

Let focus first on the two simple cases without resonances due to the loop couplings: L1A (only annihilations) and L1C (all-co-annihilations included), which are shown in the left panels in Figure \ref{fig:L1_OMH2_LRscan}. In both cases we do not have any loop induced couplings thus the cut off dependence will only enter 
in the mass spectrum. With growing cut off, for any given $m_{KK}$, the mass splittings increase (see Fig. \ref{fig:Mass_Splittings_variations_with_LamR}).
As the annihilations are mediated by the $(1)$ level particles in $t$-channels mainly (for $(1)$ quarks and $A^{(1)}$, $Z^{(1)}$), the larger the mass 
splittings the smaller the individual cross sections, thus the annihilation cross sections will be suppressed with growing $\Lambda R$. 
We show the effect in Figure \ref{fig:A1A1_annihilation_xsubsections_with_pcms_Mkk500_LR_5_10}, where we plot the annihilation cross section $A^{(1)} A^{(1)} \rightarrow SM$ for $m_{KK} = 500$ GeV as a function of the centre of mass momentum $p_{cms}$.
The two lines correspond to $\Lambda R = 5$ and $10$, thus showing that the cross section is smaller for larger cut-off.
The relic abundance depends on the mean value of $\sigma v$:  we estimate its value by averaging over a 
Boltzmann distribution of velocities for the DM
\begin{equation}
f(p)dp \sim p^2 \exp\left( \frac{p^2}{2mkT} \right)dp
\end{equation}
where $T=m/x$ with $x=25$~\footnote{We can safely assume $x$ constant. We have checked that during the relic abundance calculation 
indeed its value varies within the range of $22-26$ for all the relevant $m_{KK}$ and for all the assumed values of $\Lambda R$.}.
Then, for the mean value $\langle \sigma v \rangle$, we obtain $3.76$ pb for $\Lambda R = 5$ and 
$3.20$ pb for $\Lambda R = 10$, thus confirming that effectively the mean value $\langle \sigma v \rangle$ is larger for larger cut-off scale.
This would imply that the relic abundance is larger for larger cut-off, as confirmed in the annihilation-only case L1A.

The case with co-annihilation shows an exactly opposite behaviour.
However, a larger cut-off would imply that the individual co-annihilation cross sections decrease; furthermore, the larger mass splittings will also suppress the contribution of the co-annihilation channels.
We might then conclude that the larger the cut-off the larger the relic abundance, similarly to the annihilation only case.
However, increased mass splittings have another effect: the effective number of degrees of freedom $g_{eff}$ is also reduced.
The mass splitting increases more importantly for the leptons, while gauge bosons $W^{(1)}$ and $Z^{(1)}$ also receive a cut-off independent contribution from the $W$ and $Z$ mass.
As leptons contribute more to the degrees of freedom than to the average cross section, when increasing the cut-off the decrease in the degrees of freedom dominates and the relic abundance is reduced, as we observe in the numerical results.

Including the loop couplings does not change the trend, as it can be seen in the right panels in Figure \ref{fig:L1_OMH2_LRscan}. The only visible effect is a change in the features due to the resonant exchange of tier $(2)$ states, namely $H^{(2)}$ and $A^{(2)}$ due to the change in the resonant value of $m_{KK}$.

We summarise the numerical bounds for different values of the cut-off $\Lambda R$ for L1 and for the complete model L2 in Table 
\ref{Mkk_bounds_from_omh_nonsymm}. Note that the allowed mass values vary from about 200 GeV for the symmetric case - a value that is already excluded by the accelerator searches, up to 1 TeV - a region which is viable as not yet excluded by the LHC data. 
\begin{table}[!ht]
\begin{center}
\begin{tabular}{|c|c|c|c|c|c|c|}
\hline
 & $\Lambda R$ & 2 & 4 & 6 & 8 & 10 \\
\hline
L1 & $R_4 \gg R_5$  & 235 - 380 & 360 - 530 & 405 - 580 & 425 - 605 & 440 - 620 \\

   &$R_4 = R_5$  &  excluded & $<285$  & $<300$  & $<305$  & $<310$  \\
\hline
L2 &$R_4 \gg R_5$  & 500 - 740 & 570 - 820 & 640 - 905 & 675 - 955 & 700 - 990 \\
&$R_4 = R_5$  & 240 - 365 &295 - 455 &330 - 480 &340 - 495 &350 - 505 \\
\hline
\end{tabular}
\caption{Preferred ranges for $m_{KK}$ (in GeV) for different values of $\Lambda R$ in the asymmetric and symmetric radii cases.}
\label{Mkk_bounds_from_omh_nonsymm}
\end{center}
\end{table}

\section{Localised Higgs mass dependence}
\label{sec:locHiggs}

In this section we vary the $m_{loc}$ parameter, which is a free parameter of the model. It corresponds to the Higgs mass operator localised 
on the singular points of the orbifold. 

As it was mentioned in Section \ref{seq:results}, the main processes mediated by the resonant $H^{(2)}$ are from the 
$A^{(1)}A^{(1)}$ states. The resonant condition in Eq. \ref{eq:res_cond} depends on the loop corrections to the $A^{(1)}$ and $H^{(2)}$ masses. If we parameterise
\begin{equation}
m_{A^{(1)}}^2 = m_{KK}^2 (1+\delta_A)\,, \qquad m_{H^{(2)}}^2 = 4 m_{KK}^2 (1+\delta_H) + m_H^2 + m_{loc}^2\,,
\end{equation}
the resonant condition reads
\begin{equation}
\label{eq:Mkk_resonant_cond_for_H5}
m_{KK}^2 \leq \frac{m_h^2+m_{loc}^2}{4 \delta_A - \delta_H}.
\end{equation}
Numerically, it turns out that $\delta_A \sim 0$ and $\delta_H < 0$, therefore a resonance is possible as long as $m_{loc}^2 > - m_H^2$.
In the following we will focus on positive values of the localised mass square, which will give rise to resonance for large values of $m_{KK}$.
In Table \ref{tab:L1CLH5res} we show a list of the resonant $m_{KKres}$ and lower bound on $m_{KK}$ for various values of $m_{loc}$: we find that for $110$ GeV $< m_{loc} <  236$ GeV, the resonance always appears below the upper bound on $m_{KK}$.

\begin{table}
\begin{center}
\begin{tabular}{|c|c|c|}
\hline
$m_{loc}$ & $m_{KKres}$ [GeV]& $m_{KK} >$ [GeV]\\
\hline

0   & 267 & 0 \\

100 & 353 & 351 \\

200 & 544   & 558 \\

300 & 769& 731 \\

400 & 1005 & 886 \\

500 & 1248 & 1028 \\
\hline
\end{tabular}
\caption{Values of $m_{KKres}$ corresponding to resonance $2 m_{A^{(1)}}=m_{H^{(2)}}$ and the lower bounds for $m_{KK}$ obtained from the electroweak precision constraints (in the asymmetric case) for different values of $m_{loc}$.}
\label{tab:L1CLH5res}
\end{center}
\end{table}

Finally we study the influence of the resonant $H^{(2)}$ exchange on the relic abundance.
First we focus on the L1A case, where the effect is more visible: the only relevant annihilation process is $A^{(1)} A^{(1)} \rightarrow t \bar{t}$.
 \begin{figure}[!ht]
 \begin{center}
 \includegraphics[scale=0.85]{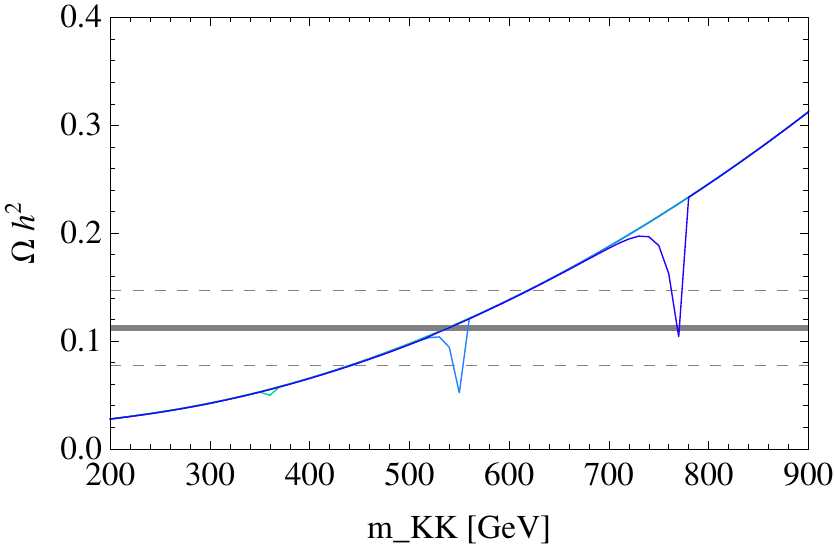}
 \includegraphics[scale=0.85]{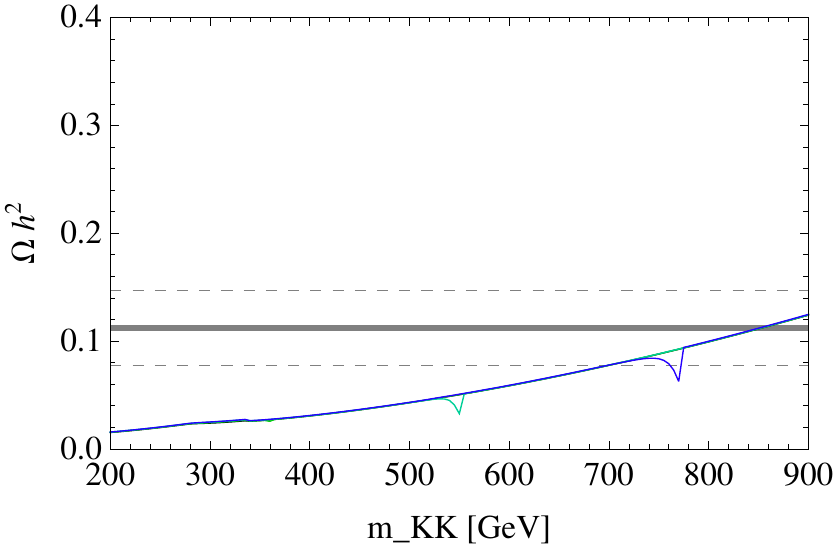}
 \caption{\footnotesize Relic abundance as a function of $m_{KK}$ with all co-annihilations and resonances included. We are particularly interested in the $H^{(2)}$ resonance as a function of $m_{loc}$. In the plots the dips correspond to the resonances of $H^{(2)}$ at, from left to right, $m_{loc} =0$, $100$, $200$, $300$ GeV. The first two minima are almost invisible, the large two dips correspond to $m_{loc} = 200,\,300$ GeV.
On the left: asymmetric L1CL. On the right: asymmetric L2CL.}
\label{fig:M6UED00ALH5res}
\end{center}
\end{figure}
In Figure \ref{fig:M6UED00ALH5res} we show the relic abundance as a function of $m_{KK}$ for various values of $m_{loc}$.
For $m_{loc} = 0$ GeV, the resonance produces a dip at low $m_{KK}$ where the relic abundance is well below the WMAP preferred region.

With increasing $m_{loc}$, the resonant dip start becoming relevant opening up extra parameter space: for instance, for $m_{loc} = 300$ GeV, we observe that a small region $760 < m_{KK} < 780$ GeV also gives the WMAP relic abundance together with the standard low $m_{KK}$ region $440 < m_{KK} < 620$ GeV. After the resonant condition is met, which gives the position of the local minimum for the relic abundance, the resonance becomes rapidly ineffective as the DM states have a mass above the KK Higgs one.
Note that the increase in the depth of the peak for increasing $m_{loc}$ is not sufficient and for large values of the localised mass the dip will not be able to touch the WMAP preferred region and no extra parameter space opens up. In the L1 model, this happens for $m_{loc} > 400$ GeV.

These features remain also in the complete model, depicted in the right panel of the Figure \ref{fig:M6UED00ALH5res}, however we see that the impact of the 
resonance is greatly suppressed. This is due to the fact that the only relevant resonant process remains $A^{(1)} A^{(1)} \to H^{(2)} \to t \bar{t}$, which is however diluted by the full set of co-annihilation processes. 

\begin{figure}[!ht]
\begin{center}
 \includegraphics[scale=0.87]{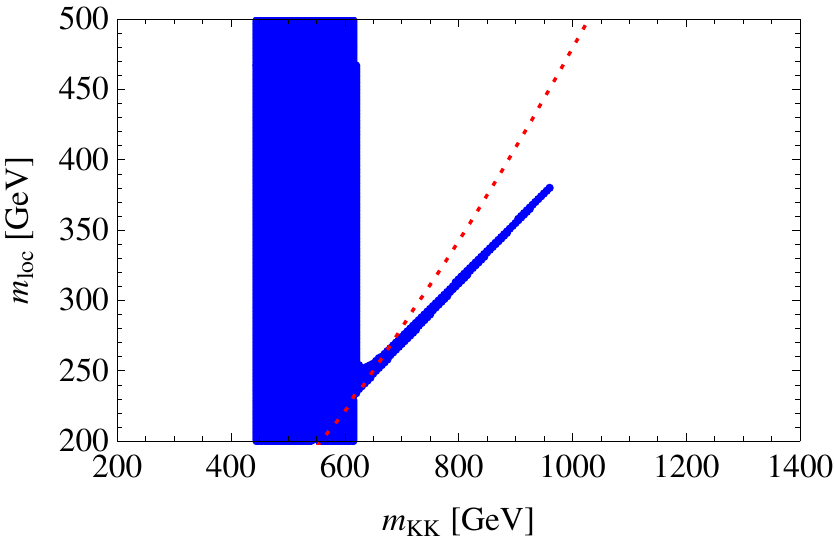}
 \includegraphics[scale=0.85]{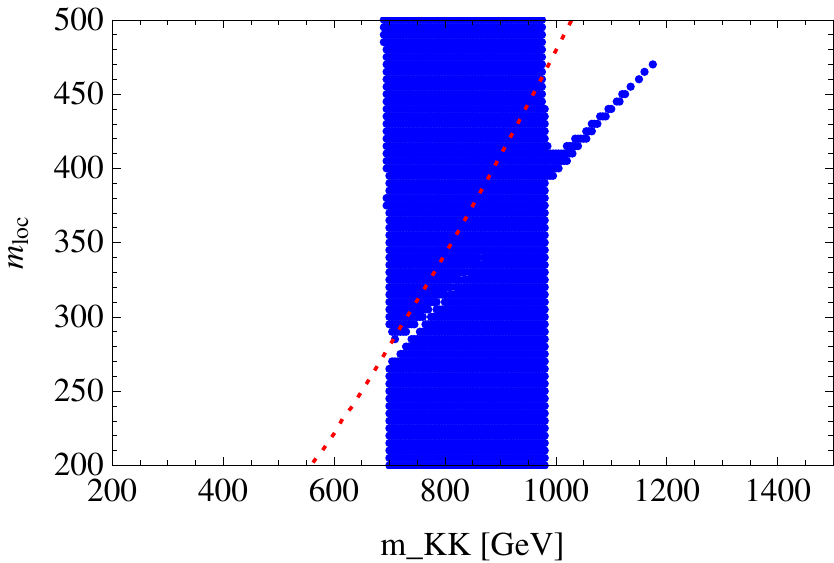}
\caption{\footnotesize We show in blue the region of $m_{KK}-m_{loc}$ preferred by WMAP in blue. The parameter space above the red line is disfavoured by the $\rho$ parameter. In the left panel: asymmetric L1 model. In the right panel: asymmetric L2 scenario.}
 \label{fig:L2CLH5res}
\end{center}
\end{figure}

To be complete we scan the parameter space $m_{KK}-m_{loc}$. The resonance has an important effect on the WMAP preferred parameter range, as it is shown in the right panel of 
figure \ref{fig:L2CLH5res}. While the usual region $640 < m_{KK} < 900$ GeV is still open, for $m_{loc} > 400$ GeV a funnel region 
opens up where the resonant $H^{(2)}$ exchange dominates.
Such region corresponds approximately to the resonant condition in Eq. \ref{eq:Mkk_resonant_cond_for_H5}, giving an allowed region around 
$m_{KK} \sim 2.5\; m_{loc}$.
Note that the funnel region closes up for $m_{loc} > 470$ GeV (corresponding to $m_{KK} \sim 1200$ GeV), where the dip in the relic 
abundance is not deep enough to touch the preferred WMAP region.
Interestingly, this implies that even the resonant funnel region admits an upper bound on $m_{KK}$.
The bound from the $\rho$ parameter, indicated by a red line in Figure \ref{fig:L2CLH5res}, shows that large values of $m_{loc}$ are 
disfavoured, however the funnel region is still in the allowed parameter space.
A similar behaviour occurs in the L1 model, depicted in the left panel of Figure \ref{fig:L2CLH5res}.

\section{Direct Detection Bounds}
\label{sec:direct}

A number of experiments are currently searching for a Cold Dark Matter. Their sensitivity is being continuously improved and the upper limits are upgraded regularly. The best upper limit for the WIMP-proton spin independent cross section has been recently obtained by Xenon, $\sigma^{SI}_{p\chi} < 3.4\times 10^{-8}$ pb for a 55 GeV WIMP and CDMS $\sigma^{SI}_{p\chi} < 3.8\times 10^{-8}$ pb for a 70 GeV WIMP. 

In the extra-dimensional models the potential dark matter candidate is usually a KK gauge boson, but it is also possible that a KK scalar, a KK graviton or a KK neutrino is the Lightest KK particle. In the minimal UED model (mUED) the dark matter candidate is the $U(1)$ gauge boson $B^{1}$\footnote{More precisely the Dark Matter Candidate is the level 1 photon $A^{(1)}$ - a linear superposition of $B^{(1)}$ and $W^{(1)}_3$ - but the mixing angle can be safely neglected.}. In our six-dimensional model the dark matter candidate is the first level photon $A^{(1)}$ which is a scalar particle. Thus its interaction properties will be different from the 5D UED models where the first level photon is a vector particle, as spin-dependent interactions are absent. Therefore the direct detection will be mediated only by spin-independent interactions. The case where the direct detection signal is mediated only by the SI interaction is much less discriminating with respect to which model is being detected. The only information that can be used is the total cross section and the mass of the DM particle.

\begin{figure}[!ht]
\begin{center}
\includegraphics[scale=0.8]{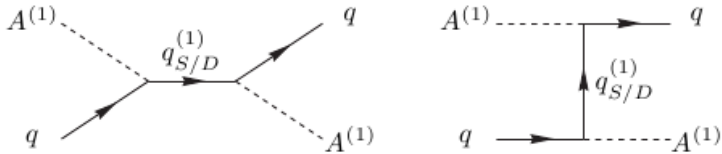}
\caption{\footnotesize Feynman diagrams for direct detection $A^{(1)} q \rightarrow A^{(1)} q$ via $q^{(1)}_{S/D}$ in $s$ and $t$ channel; $q$ stands for the quarks present in the nucleon, thus mainly up and down.}
\label{Direct_detection_feynman_diagrams}
\end{center}
\end{figure}

We compute the direct detection signal using MicrOMEGAs v2.4.1. The input model is as defined in previous subsection.
The direct detection signal is mediated by the processes showed in Figure \ref{Direct_detection_feynman_diagrams}.

The interactions of $A^{(1)}$ with all the SM quark are mediated by the level one quarks $q^{(1)}_{D/S}$ in $s$ and $t$ channel. Interactions with heavy quarks $c$, $b$, $t$ are also mediated by the SM Higgs boson $h$ in $t$-channel. The Yukawa couplings with light quarks can be safely neglected  as they are proportional to the mass of the quark. 

The model has two free parameters: $m_{loc}$ and $\Lambda R$.
The latter only influences the KK Higgs masses and it can only enter via a $H^{(2)}$ $t$-channel exchange, which is negligible as it is suppressed by one loop compared to the tree level diagrams in Figure  \ref{Direct_detection_feynman_diagrams}.
On the other hand, the result depends crucially on the cut-off $\Lambda R$: in fact, increasing the cut-off will increase the mass of the tier $(1)$ quarks and thus suppress the scattering.
The effect is important due to the closeness between the mass of the DM candidate $A^{(1)}$ and the heavy quarks.
In our numerical study, we vary the $\Lambda R$ parameter in the range
\begin{equation}
2 < \Lambda R < 10 \,,
\end{equation}

\begin{figure}
\begin{center}
\includegraphics[scale=1.2]{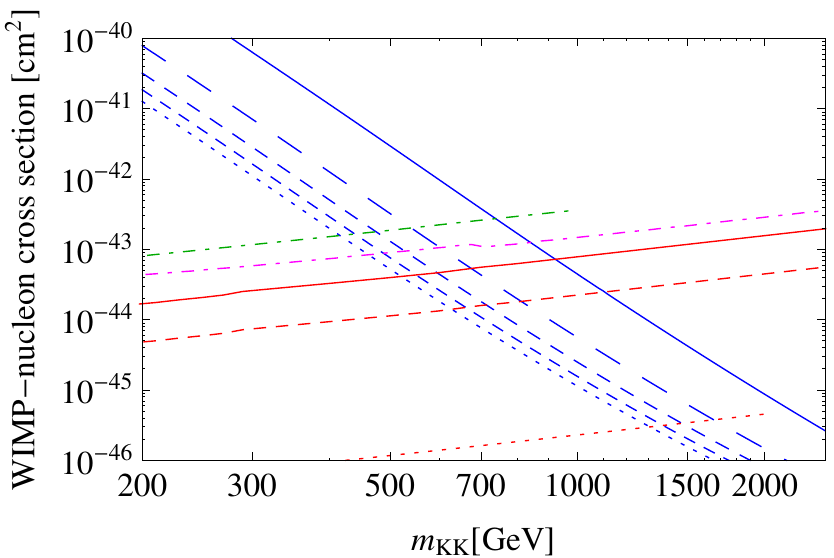}
\caption{\footnotesize Direct Detection bounds on $m_{KK}$ for the RP$^2$ for $\Lambda R = 2,4,6,8,10$. The WIMP - proton cross 
sections in blue correspond to the 5 values of the cut-off: from blue solid line ($\Lambda R = 2$) down to blue dotted line ($\Lambda R = 10$).  In red we show the Xenon limit: solid - 
2011, dashed - 2012, dotted - 2017. In dot-dashed green/magenta the limits from Zepelin/CDMS+EDELWEISS.}
\label{fig_M6UED_SIcrossSection_and_limits_LR3_5_10_plot}
\end{center}
\end{figure}

\begin{table}
\begin{center}
\begin{tabular}{|c||c|c|c|c|c||c|}
\hline
	Experiment                            & Zepelin & Edelweiss & Xenon2011  & Xenon2012 & Xenon2017 & $\Lambda R$ \\ 
	   &   &+CDMS &  &  &  &  \\
\hline
	$10^{-44}\sigma\textrm{[cm]}^2$       & 27.089  & 12.589         & 6.422      & 2.061     & 0.05      & 2           \\
	$m_{KK}$[GeV]                         & 740     & 840            & 940        & 1140      & 2200      &             \\
	\hline
          $10^{-44}\sigma\textrm{[cm]}^2$       & 185     & 97.285         & 45.308     & 14.555    & 0.251     & 4           \\
	$m_{KK}$[GeV]                         & 540     & 600            & 680        & 820       & 1650      &             \\
	\hline
	$10^{-44}\sigma\textrm{[cm]}^2$       & 435.537 & 209.142        & 97.285     & 32.004    & 0.498     & 6           \\
	$m_{KK}$[GeV]                         & 470     & 530            & 600        & 720       & 1460      &             \\
	\hline
	$10^{-44}\sigma\textrm{[cm]}^2$       & 643.533 & 337.704        & 166.731    & 54.356    & 0.685     & 8           \\
	$m_{KK}$[GeV]                         & 440     & 490            & 550        & 660       & 1380      &             \\
	\hline
	$10^{-44}\sigma\textrm{[cm]}^2$       & 14.718  & 7.615          & 4.165     & 1.38922   & 0.028     & 10           \\
	$m_{KK}$[GeV]                         & 415     & 465            & 520       & 630       & 1315      &               \\
	\hline
	\end{tabular}
	\caption{\footnotesize Upper bounds on $m_{KK}$ mass from different direct detection experiments.}
\label{tab:M6UED_DD_bounds_LRvariable}
	\end{center}
	\end{table}

\begin{figure}
\begin{center}
\includegraphics[scale=0.8]{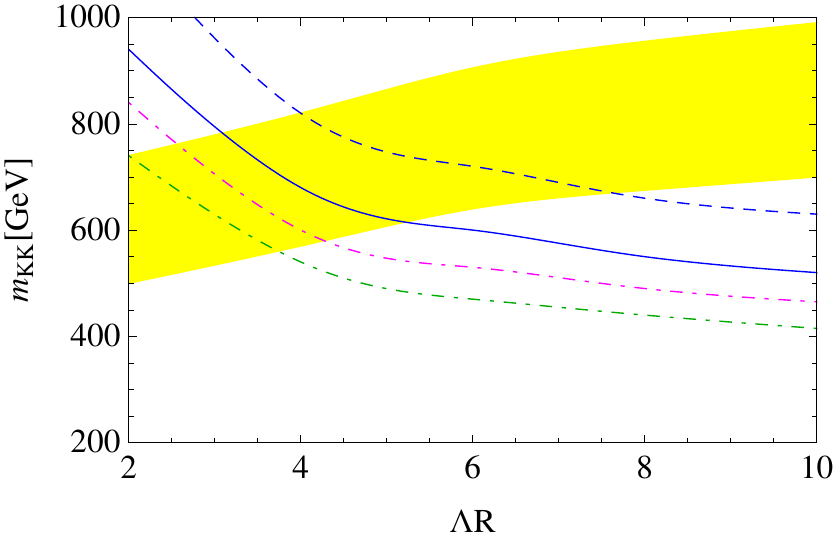}~~
\includegraphics[scale=0.8]{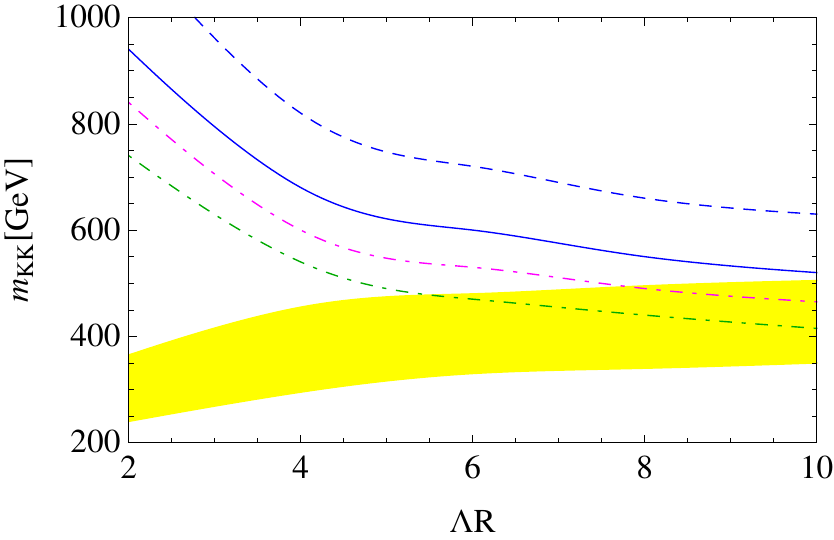}
\caption{\footnotesize In yellow we show the WMAP preferred region for $m_{KK}$ as a function of the cut-off scale $\Lambda R$ for asymmetric (left) and symmetric (right) radii. The lines delimit the lower bounds from direct detection experiments: Zepelin (dot-dashed green), CDMS+EDELWEISS (green dot-dashed), Xenon 2011 (blue solid), Xenon 2012 projection (blue dashed), Xenon 2017 projection (blue dotted). The allowed $m_{KK}$ regions are above the direct detection lines.}
\label{fig_crossbounds}
\end{center}
\end{figure}

The results are shown in Figure \ref{fig_M6UED_SIcrossSection_and_limits_LR3_5_10_plot} where we plot the spin-independent cross section as a function of $m_{KK}$ for five choices of $\Lambda R$ (in blue) as well as various bounds from experiences: Zepelin \cite{Akimov:2011tj}, Edelweiss + CDMS combined data \cite{Ahmed:2011gh}, Xenon limits from 2011 \cite{Aprile:2011hi} and 2012 \cite{Aprile:2011dd} as well as projection limits in Xenon for 2017 extracted from DMTools. The numerical values for $\Lambda R = 2$, $4$, $6$, $8$, and $10$ are given in Table \ref{tab:M6UED_DD_bounds_LRvariable}.
The best bound is coming from Xenon 2011, and it ranges from $m_{KK} > 520$ to $760$ GeV, increasing for smaller cut-offs.
The bound is also independent on the radii, being the same foe asymmetric and symmetric ones.
It is interesting to compare the bounds in Table \ref{tab:M6UED_DD_bounds_LRvariable} with the preferred WMAP ranges in Table \ref{Mkk_bounds_from_omh_nonsymm}: we immediately see that the WMAP preferred ranges move to higher values of $m_{KK}$ for larger cut-offs, while the bounds from direct detection decrease.
Numerically, the symmetric case seems to be completely excluded, while low cut-off values $\Lambda R \lesssim 3$ are excluded for asymmetric radii.
This is clear from Figure \ref{fig_crossbounds}, where we show in yellow the region preferred by WMPA in the $\Lambda R$--$m_{KK}$ region together with the bounds from direct detection experiments: the best bound is given by Xenon100 (solid blue line).
It is also interesting to compare the Direct Detection bound with bounds from accelerators (LHC): the strongest bound should come from dilepton resonances from the decays of the even tiers $(2)$.
An estimate \cite{RP2spectrum} shows that the bound, after the analysis of the 2011 data at 7 TeV, is $m_{KK} < 575$ GeV in the asymmetric case, and $m_{KK} < 440$ GeV in the symmetric one~\footnote{Indirect bounds on $m_{KK}$ can be extracted from the Higgs physics \cite{Nishiwaki:2011gk,Kakuda:2012px}, and may be important.}.
The bounds have been computed for $\Lambda R = 10$, however they should bear a minor dependence on its precise value and this be near constant in the region under investigation.
Direct detection bounds are therefore competitive with collider ones, and in particular dominate for smaller mass splitting (small $\Lambda R$) and in the symmetric case.

One way out of this conclusion is the funnel region opened by the $H^{(2)}$ resonance, which allows for large $m_{KK}$.
The strong direct detection bounds can also be compared to the bounds on other UED models in 2 dimensions, like the chiral square, where smaller cross sections are obtained \cite{Dobrescu:2007ec}.

\section{Conclusion}
\label{sec:conclusion}

We have studied the dark matter relic abundance on the Universal Real Projective Plane model. This model is an extra dimensional 
model with two extra dimensions compactified on a twisted bottle, or real projective plane. This peculiar geometry brings an extra argument for the stability 
of dark matter,  which is not an imposed parity but just a remnant of the 6-dimensional Lorentz invariance partially broken by the 
compactification. One of the main features of this model is the highly degenerate spectrum in each Kaluza-Klein tier, due to the smaller loop corrections compared to other UED models. This in turn implies 
that co-annihilation effects and higher modes are important in the calculation of the relic abundance. We have performed a detailed 
analytic and numerical study which allows to obtain a range of KK-masses in which the model is consistent with the relic abundance by 
providing a good candidate for the stable dark matter particle. 
Our results can be compared to calculations performed for other UED models \cite{Kong:2005hn,Burnell:2005hm,Kakizaki:2006dz,Belanger:2010yx}.

Our complete computation of the relic density of dark matter in the Universal Real Projective Plane model includes all effects of 
tier $(1)$ and tier $(2)$ states together with the precise spectra calculated at one-loop level, thus allows to understand in detail the role 
of the different ingredients active in this model: annihilation versus co-annihilation, 
different particles contributions for DM observables, effect of the cut--off and of localised Higgs mass terms, perspective of direct versus indirect 
bounds, implication for the LHC searches.
We find that the preferred range for $m_{KK}$ is $700 < m_{KK} < 990$ GeV for the asymmetric case and $350 < m_{KK} < 505$ GeV for degenerate radii, in both cases for maximal cut-off $\Lambda R = 10$. Decreasing the cut-off, the preferred ranges are reduced.
Such ranges are well into the reach of the LHC.
We also find that the tier $(2)$ Higgs resonance open a funnel region in the parameter space that allows to increase $m_{KK}$ up to $\sim 1200$ GeV.
We also computed the SI direct detection cross section for the scalar Dark matter candidate, and we found values close to the experimental bounds: in fact, we found that the 2011 Xenon100 results already exclude the degenerate radii case, and severely constraints the low cut-off case for asymmetric radii.
We expect that by the end of 2017, Xenon may be able to exclude the full parameter space.
Direct detection bounds are also competitive with LHC bounds, even though a complete study of the Universal RP$^2$ at the hadron collider has not been completed.

Dark Matter in extra dimensions is still a very plausible candidate to explain the matter density in the Universe, and it offers the possibility of an exact symmetry deriving from the properties of the compact space.
Furthermore, the mass scales required by the WMAP data are in the range presently being explored at the LHC.
We explored a particular realisation of this idea on a twisted bottle, while many other spaces and topologies are still unexplored.

\subsection*{Acknowledgements}
We would like to thank A. Pukhov for his help with MicrOMEGAs. A.A. acknowledges partial support from the European Union FP7 ITN INVISIBLES (Marie Curie Actions, PITN-GA-2011-289442).
G.C. acknowledge the hospitality of King's College London during the completion of this work.


\end{document}